\documentclass[aps,pre,eqsecnum,twocolumn,showpacs]{revtex4}
\usepackage{amsmath,bm,epsfig}

\def\Fbox#1{\vskip1ex\hbox to 8.5cm{\hfil\fboxsep0.3cm\fbox{%
  \parbox{8.0cm}{#1}}\hfil}\vskip1ex\noindent}  
\let \nn  \nonumber

\newcommand{\br}{\\ \nn}
\def\r{\bm r}
\def\R{\bm R}

\def\vt {v\Sb T }
\def\Wd {W^\ddag} \def\Vd {V^\ddag}
\def\Sd {S^\ddag}
\def\vt {v\Sb T ^\dag}\def\vd {v^\ddag}
\def\yp {y^+}
\def\yd {y^\ddag}
\def\Wp {W^+} \def\Vp{V^+}
 \def\gd {\~{\gamma}^\ddag} \def\Gd {\Gamma^\ddag}
 \def \vK {von-K\'arm\'an~}

\let\*\cdot
\def\<{\left\langle} \def\>{\right\rangle} \def\({\left(} \def\){\right)}
\let\p\partial \let\~\widetilde \let\^\widehat \def\ort#1{\^{\bf{#1}}}
 \def\x{\ort x} \def\y{\ort y}
  \def\1{\bm1} 
 
\newcommand{\B}[1]{{\bm{#1}}}
\def\BE{\begin{equation}}\def\EE{\end{equation}}
\def\BEA{\begin{eqnarray}}\def\EEA{\end{eqnarray}}
\def\BSE{\begin{subequations}}\def\ESE{\end{subequations}}

\renewcommand{\sb}[1]{_{\text {#1}}}  
\renewcommand{\sp}[1]{^{\text {#1}}}  
\newcommand{\Sb}[1]{_{_{\text {#1}}}} 

\newcommand{\Ref}[1]{(\ref{#1})}
\newcommand{\REF}[1]{Eq.~(\ref{#1})}

\renewcommand{\a}{\alpha}
\newcommand{\g}{\gamma}
\newcommand{\G} {\Gamma}\renewcommand{\d}{\delta}
\newcommand{\D}{\Delta}\newcommand{\e}{\epsilon}

\let \= \equiv

 \def\r{\bm r} \let\p\partial 
 \let\*\cdot \def\sb#1{_{\rm{#1}}}
 \def\R{\mathcal R}
 \def\({\left(} \def\){\right)}
 \def \[ {\left [} \def \] {\right ]}
   
  \def\1{\bm\delta}  \let\^\widehat \def\REF#1{Eq.~\Ref{#1}}
   \def\Ref#1{(\ref{#1})}  \def\<{\left\langle}
   \def\>{\right\rangle}  \def\x{\ort x}
   \def\ort#1{\^{\bf{#1}}} \def\y{\ort y} 
     \def\Sb#1{_{\scriptscriptstyle\rm{#1}}}
     \let\~\widetilde

\begin{document}

\title{Phenomenology of Wall Bounded Newtonian Turbulence}
\author{Victor S. L'vov} \email{Victor.Lvov@Weizmann.ac.il}
\homepage{http://lvov.weizmann.ac.il}
\author{Anna Pomyalov}
\email{Anna.Pomyalov@Weizmann.ac.il} \author{Itamar Procaccia}
\email{Itamar Procaccia@Weizmann.ac.il}
\homepage{http://www.weizmann.ac.il/chemphys/cfprocac/home.html}
\affiliation{Department of
Chemical Physics, The Weizmann Institute of Science, Rehovot
76100, Israel }
 \author{Sergej S. Zilitinkevich}\email{Sergej.Zilitinkevich@fmi.fi}
\affiliation{  Division of Atmospheric Sciences Department of Physical 
Sciences
University of Helsinki, 00101 Helsinki, Finland, \\  {\rm and}   Nansen Environmental and Remote Sensing Centre / Bjerknes Centre for Climate Research, Bergen, Norway }

\begin{abstract} 
We construct  a simple analytic  model for wall-bounded turbulence, containing only four adjustable parameters.  
Two of these parameters characterize the viscous dissipation of the components of the Reynolds stress-tensor  and other two parameters characterize their nonlinear relaxation.  The model offers an analytic description of the profiles of the mean velocity and the correlation functions of velocity fluctuations in the entire boundary region, from the viscous  sub-layer, through the buffer layer and further into the log-layer.  As a first approximation, we employ the traditional return-to-isotropy hypothesis, which yields a very simple distribution of the turbulent kinetic energy between the velocity components in the log-layer: the streamwise component contains a half of the total energy whereas the wall-normal  and the cross-stream components contain a quarter each. In addition, the  model predicts a very simple relation between the von-K\'arm\'an slope $\kappa $ and the turbulent velocity in the log-law region $v^+$ (in wall units): $v^+=6 \kappa$. These predictions are in excellent agreement with DNS data and with recent laboratory experiments. \end{abstract}
\pacs{43.37.+q,47.32.Cc, 67.40.Vs, 67.57.Fg}
\date{\today}
  
\maketitle
\section*{\label{s:intro}Introduction}

The tremendous amount of work devoted to understanding the apparent experimental {\em deviations} from the classical phenomenology of homogeneous and isotropic turbulence~\cite{MY,Fri} tends to obscure the fact that in many respects this phenomenology is almost right on the mark. Starting with the basic ideas of Richardson and Kolmogorov, and continuing with a large number of ingenious closures, one can offer a reasonable set of predictions regarding the statistical properties of the highly complex phenomenon of homogenous and isotropic turbulence. Thus one predicts the range of scales for which viscous effects are negligible (the so called ``inertial range" of turbulence), the cross over scale below which dissipative effects are crucial (also known as the ``Kolmogorov scale"), the exact form of the third order structure function $S_3(R)$ (third moment of the longitudinal velocity difference across a scale $R$),  including numerical pre-factors, and an approximate form of structure function of other orders $S_n(R) $ (predicted to scale like $R^{n/3}$ but showing deviations in the scaling exponents which grow with the order, giving rise to much of the theoretical work alluded to above). In particular much effort was devoted to calculate the so called ``Kolmogorov constant" $C_2$ which is the pre-factor of the second order structure function, with closure approximations (see, e.g. Refs.~\cite{66Kra,86YO}) coming reasonably close to its experimental estimate.  Notwithstanding the deviations from the classical phenomenology, one can state that it provides a reasonable first order estimate on many non-trivial aspects of homogeneous and isotropic turbulence.  In contrast, the phenomenological theory of wall bounded turbulence is  less advanced. In reality most turbulent flows are bounded by one or more solid surfaces, making wall bounded turbulence a problem of paramount importance. Evidently, a huge amount of literature had dealt with problem, with much ingenuity and considerable success~\cite{00Pope}. In particular one refers to von-K\'arm\'an's log-law of the wall which describes the profile of the mean velocity as a function of the distance from the wall. It appears however that the literature lacks an analytically tractable model of wall bounded flows whose predictions can be trusted at a level comparable to the phenomenological theory of homogeneous turbulence.

 In this paper we attempt to reduce this gap and offer as simple as possible but still realistic model (a "minimal model") for the viscous-to-turbulent flow in the entire region from the very surface through    the logarithmic boundary layer (hereafter, \emph{log-layer}) up to the upper boundary of   outer   turbulent region.  Our final goal is to create clear physical grounds for improved description of the flow-surface interactions in numerical fluid-mechanics models (both engineering and geophysical) where the viscous and the buffer layers cannot be resoled and should be parameterized. To attain these ends we need to obtain analytical solutions (numerical solution would be of no use), which calls for simplification of the governing equations. Accordingly, our strategy is a pragmatic, task-dependent simplification and restrictions. In particular we concentrate on  descriptions of  the profile of mean flow and the statistics of turbulence on the level of simultaneous, one-point, second-order velocity correlation functions.
 In other words, the objects that we are after are the entries of the Reynolds-stress tensor {\em as a function of the distance from the wall}.  The model will be presented for plain geometry; this geometry is relevant for a wide variety of turbulent flows, like channel and plain  Couette flows,  fluid flows over inclined planes under gravity (modelling river flows), atmospheric turbulent boundary layers over flat planes and, in the limit of large Reynolds numbers, many other turbulent flows, including pipe, circular Couette flows, \emph{etc}. 

Suggested in this paper phenomenological theory of wall-bounded flows  is based on standard ideas~\cite{00Pope}; nevertheless we develop the theory slightly further than anything that exists currently in the literature.  In our study we will stress analytical  tractability; in other words, we will introduce approximations in order to achieve a model whose  properties and predictions can be understood without resort to numerical calculations. Nevertheless  we will show that the model appears very dependable in the sense that its predictions check very well in comparison to direct numerical simulations, including some rather non-trivial predictions that  are corroborated only by very recent simulations and experiments (which only now reach the sufficient accuracy and high Reynolds numbers).  

We should notice, that  considering the mean velocity and the second-order statistics in these (neutrally stratified) flows, we neglect some mechanisms and features although present but not essential in the problem under consideration. However, proceeding further, in particular, to account for the density or temperature stratification (out of the scope of this paper), we quite probably will be forced to rule out of some simplifications acceptable in the first task and to account, for example, for a spacial energy flux and even for coherent structures.  
 
In Sec.~\ref{s:min-mod} we formulate a model which is a version of  the ``algebraic Reynolds-stress model"~\cite{00Pope}. In Sec.~\ref{ss:mech} we  introduce notations and recall the equations describing the mechanical balance; in Sec.~\ref{ss:energy} we state the assumptions and detail the approximations used in the context of  the balance equations for the components of the Reynolds stress tensor $W_{ij}$. The result of these considerations is a set of 5 equations for the mean shear $S$ and $W_{ij}$ which is described in Sec.~\ref{ss:sum}. For actual calculations this set of equations is still too rich since it contains  12 adjustable parameter. Eight  of these parameters control the nonlinear behavior of the system in the outer layer and four additional parameters govern the energy dissipation in the viscous sub-layer. Clearly, further reduction of the model is called  for. This is accomplished in Sec.~\ref{s:free}.  First, in Sec.~\ref{ss:gen-an} we consider the full 12 parametrical solution of the model, and present a comparison with experimental observations in Sec.~\ref{ss:4-par}.  This comparison indicates that an adequate description of the entire turbulent boundary layer phenomenology can be achieved with only four parameters instead of twelve.  We refer to the four-parameter model as the ``minimal model". In Sec.~\ref{ss:turb} we reap the benefit of the minimal model: we find simple and physically transparent Eqs.~(\ref{turb1})  for the profiles of the Reynolds stress tensor $W_{ij}(y)$ and the mean   shear $S(y)$ ($y$ is the distance from  the wall) in terms of the root-mean-square   turbulent velocity $v\=\sqrt{\mathstrut W_{ii}}$. Unfortunately, the equation for the $v(y)$ profile is quite cumbersome and cannot be solved analytically.  Nevertheless we employ an effective iteration procedure that allows reaching highly accurate solutions with one or at most two iteration steps.

Section~\ref{s:exp} is devoted to a comparison of the predictions of the minimal model with results of experiments and direct numerical simulations . In particular, in Sec.~\ref{ss:meanV} we show that the model describes the mean velocity profile in a channel flow with $1\%$-accuracy almost everywhere.  Only in the core the model fails to describe so-called ``velocity defect" (the upward deviation
from the log-law)  which is observed near the mid-channel (of about 5-6
units in $V^+$, independent on Reynolds number).  For our purposes this mismatch in not essential. In Sec.~\ref{ss:Kin}  we show that  the minimal model provides a good qualitative description of kinetic energy profile, including position, amplitude and width of the peak of the kinetic energy in the buffer layer.  In Sec.~\ref{ss:Wxy-Wii}  we show that with the same set of four   parameters the model offers also a good qualitative description of the Reynolds stress profiles and the profiles of  ``partial" kinetic energies (in the streamwise, wall-normal and cross-stream directions)  almost  in the entire channel.  The final  Sec.~\ref{s:sum} presents a short summary of our results, including a discussion of the limitations of the minimal model. Possible improvements of the suggested model will have to start by addressing
these limitations.

\section{\label{s:min-mod}Formulation of the model}
Our starting point is the standard Reynolds decomposion~\cite{00Pope} of
the fluid velocity $\B U(\r, t)$ into  its average (over time) $\B V $
and  the fluctuating components $\B u$. In wall-bounded planar geometry  
the mean
velocity is oriented in the (stream-wise) $\x$ direction,
depending on the vertical (wall-normal) coordinate $y$ only:
\begin{equation}\label{dec-a}
\B U(\B r,t) = \B  V( y) + \B u(\B r,t) \ , \quad \B V(y) \equiv
\langle \B U(\B r,t) \rangle = \x \, V(y)\ .
\end{equation}
The mean velocity and the fluctuating parts are used to construct the 
objects of the theory which are the  components of the Reynolds stress 
tensor $\B W(y)$ and the mean shear:
\BE \label{defs}
W_{ij}(y) \= \< u_i u_j \>\,, \quad S(y)\=\frac{d V(y)}{d y}\ .
\EE
We note that in previous applications \cite{04LPPT,04DCLPPT,04BLPT,05LPPT,05LPPTa,04BDLPT,04LPT,05BCLLP,05BDLP} we have employed a model in which only the trace of $\B W(y)$ and its $xy$ component were kept in a simplified description. For the present purposes we consider all the component of this tensor, paying a price of having more equations to balance, but reaping the benefit of a significantly improved phenomenology. We discuss now the equations relating these variables to each other.
\subsection{\label{ss:mech} Equation for the mechanical balance}
The first equation relates the Reynolds stress $W_{xy}$ to the mean shear; it describes the balance of the flux of mechanical momentum, it follows as an exact result from Navier-Stokes equations  and has the familiar form: \BSE \begin{equation}\label{Mflux} -W_{xy}(y)+\nu_0 S(y) =P(y) \ . \end{equation} 
  $W_{xy}$ on the left hand side (LHS) is the turbulent  (reversible) contribution to the momentum flux whereas $\nu_0 S(y)$ is the  viscous (dissipative) contribution to the momentum flux. The RHS is the momentum flux, which may have different origin. For example, in a channel or pipe flow $P(y)$ is generated by the pressure head. In the  channel flow with  the pressure gradient $p'=-d p/d x$,    
 \begin{equation}\label{ProdM} P(y)=p' (L-y)\,,
\end{equation}\ESE    %
where $L$ is the half width of the channel.  In the pipe flow $P(y)$ is given by the same Eq.~\Ref{ProdM} with $L$ being a half of the pipe radius. In a water flow over incline in the gravity field  $p'$ should be replaced by $g \sin a$, where $g$ is the gravity acceleration and $\a$ is the inclination angle.        For    Re$_\lambda\gg 1$,  near the wall one can neglect the $y$ dependence of $P(y)$, replacing $P(y)$ by its value at the wall: $P(y)\Rightarrow P_0\equiv P(0)$.  Here the   so-called ``wall-based" Reynolds   number Re$_\lambda$ is  defined by:
 \begin{equation}
  \mbox{Re}_\lambda  \equiv 
  \frac{L\sqrt{\mathstrut P(0)}}{\nu_0}\ . 
  \label{Rl}
\end{equation}     
\subsection{\label{ss:energy} Balance of the Reynolds  tensor}
The next set of equations relates the various components of
 the Reynolds  tensor,  $W_{ij}(y)$ defined by~\REF{defs}. In contrast
 to Eq.~(\ref{Mflux}) this set of equations is only partially exact. We 
need to
 model some of the terms, as explained below. We start from the 
Navier-Stokes equations
 and write the following set of equations:
 \begin{subequations}\label{bal-K}
\BE\label{bal-Ka}
\frac{d W_{ij}}{d\, t}+\epsilon_{ij}+  I_{ij}=
- S(W_{iy}\delta_{jx}+W_{jy}\delta_{ix}) \ . 
\EE
The RHS of these equations is exact, describing the  production term
in the equations for $W_{ix}=W_{xi}$ which is caused by the existence of 
a mean shear.
On the LHS of \REF{bal-Ka} 
\BE\label{dis-en} 
\epsilon_{ij}=2\nu_0
\<\frac{\p u_i}{\p x_k}\frac{\p u_j}{\p x_k}\> 
\EE
is the exact term  presenting  the viscous energy dissipation. The problem is that $\epsilon_{ij}$ involve new object,  which requires evaluation via $W_{ij}$. This can be easily done  in regions  where the velocity field is rather smooth, and in particular in the viscous sub-layer, the velocity gradient exists and thus the spatial derivatives in \REF{dis-en} are estimated using a characteristic length which is the distance from the wall $y$. In order to write equalities we employ the dimensionless constants $a_{ij}\simeq 1$:
\BE\label{est1v}
 \epsilon_{ij}\Rightarrow \epsilon_{ij}\sp {vis} =  \gamma_{ij}\sp {vis} W_{ij}\,, \quad 
 \gamma_{ij}\sp {vis} \simeq  \nu_0\Big(\frac{a_{ij}}{y }\Big)
 ^2\  .
 \EE\ESE %
In general the constants $a_{ij}$ are different for every $i,j$.   

In the  buffer sublayer and in the log-layer the energy
cascades down the scales until it dissipates at the Kolmogorov
(inner) scale that is much smaller than the distance
$y$ from the wall.  Therefore the main  contribution to the dissipation   $\epsilon_{ij}$  from all scales smaller than
$y$ is due to the energy flux, i.e. has a \emph{nonlinear character}. Due to the asymptotical isotropy of fine-scale turbulence, the nonlinear contribution should be diagonal in $i$, $j$ (see, e.g.~\cite{00Pope}):
\BSE\label{flux}\BE\label{flux1}
\epsilon_{ij}\Rightarrow \epsilon_{ij}\sp {nl} = \gamma \,\frac W3 \,   \delta_{ij}\,, \quad W\= \mbox{Tr}\{\B W\}\,,
\EE
where  prefactor $\frac 13$ is introduced   to simplify equations below.    
The characteristic ``nonlinear flux frequency"  
$\gamma\sp {nl}$,  can be estimated using a standard  Kolmogorov-41 
dimensional analysis: 
\BE\label{est1v}
\gamma (y)= \frac {b }{y}\sqrt {W(y)}\ , \quad  \ ,
\EE\ESE%
again with some  constants $b \sim 1$. The  ``outer scale" of turbulence is estimated in Eq.~({\ref{est1v}) by the only available characteristic length, $y$, the distance to the wall.

 As one sees from~\REF{flux}, the dissipation of particular component of the Reynolds-stress tensor, say $W_{xx}$,  depends not only on $W_{xx}$ itself, but also on other components, $W_{yy}$ and $W_{zz}$ in our case. It means that $ \epsilon_{ij}$, given by \REF{flux1} leads, in the framework of \REF{bal-K} not only to the dissipation of total energy, but also to its redistribution between different components of $W_{ii}$.  In order to separate these effects we   divide $\epsilon_{ij}$ into two parts
as follows:
\BSE \BEA\label{sep1}
\e_{ij}\sp {nl }&=&\e _{ij}\sp {nl,1}+\e_{ij}\sp {nl,2}\,,\\ \label{sep11}
\e _{ij}\sp {nl,1}&=& \g    W_{ij}\delta_{ij}\,,  \\ \label{sep12}
\e_{ij}\sp {nl,2}&=& - \g  \( W_{ij} - \delta_{ij} \frac W  3  \)\delta_{ij}\ .
 \EEA  \ESE
 Clearly, $\e _{ij}\sp {nl,1}$ describes the   damping  of each component $W_{ii}$ separately, without changing of their ratios, while the traceless part, $\e _{ij}\sp {nl,1}$,  does not contribute to the dissipation of total energy and leads only to redistribution of energy between components of the Reynolds-stress tensor.    This contribution we will include into the ``return to isotropy" term~\Ref{isotr},  that will be discussed below. 

 Actually, we presented $\e_{ij}$ as the sum
 \BE \label{div}
 \epsilon_{ij}=\e_{ij}\sp {dis}+ \e_{ij}\sp{nl,2}\,,
 \EE
 in which for the total energy dissipation is responsible only first term in the RHS.   In the buffer layer both contributions to $\e_{ij}\sp {dis}$,     the viscous dissipation, $\e_{ij}\sp {vis}$, and the nonlinear one,  $\e _{ij}\sp {nl,1}$  are important.  Their relative 
role  depending on the turbulent statistics. We will employ two simple
interpolation formulas which lead to two versions of the minimal model:
 \BEA\label{en-dis} 
 \e_{ij}\sp {dis}&=&  \G_{ij} W_{ij}\,, \\ \label{int-sum}
 \Gamma_{ij}(y) &=&  \gamma_{ij }\sp {vis}(y) +\gamma(y)\, \delta_{ij}\,, \qquad 
\mbox{``sum"}\,,\\
\label{int-root}
 \Gamma_{ij}(y) &=&  \sqrt{\gamma_{ij,\rm vis}^2(y) +\gamma^2(y)\, \delta_{ij}}\,, 
\quad \mbox{``root"}\  .
\EEA
The versions of the resulting model will be referred to as the ``sum" and the  ``root" versions correspondingly. A-priori there is no reason to prefer one or the other, and we leave the choice for later, after the comparisons with the data.

The  term $I_{ij}$  in Eq. (\ref{bal-Ka}) is caused by   the pressure-strain correlations:
\BE I_{ij}=-\frac1\rho_0\<p\(\frac{\p u_i}{\p x_j}+ \frac{\p u_j}{\p x_i}\)\>\,,
\EE
and is known in the literature as the ``Return to Isotropy" ~\cite{00Pope}. 
Due to incompressibility  constraint  $I_{ij}$ is traceless tensor and therefore does not contribute to the total energy balance, leading only to redistribution of  partial kinetic energy between different vectorial components. Also,  this term does not exist in isotropic turbulence where $W_{ij}=\frac13 W \delta_{ij}$. We adopt the simplest linear Rota approximation for the  ``Return to Isotropy" term ~\cite{00Pope}, using yet another  different characteristic frequencies $\overline{ \gamma} _{ij}$, estimated as follows:
\begin{subequations}\label{rotta1}
\BEA\label{isotr1} 
I_{ij}&=&  \overline{\gamma} _{ij}\(3\,W_{ij}-\delta_{ij}W\)\,, \\
\label{est51} 
 \overline{\gamma} _{ij}(y) &\=& \overline{ b }_{ij} \frac{\sqrt{W(y)}}{y }\,,\quad  
\overline{ b }_{ij}\simeq 1\ .
\EEA\end{subequations}
One sees that $I_{ij}$ has precisely the same structure as $\e_{ij}\sp{nl\,,2}$, introduced by \REF{sep12}. Therefore it is convenient to treat these contributions together, introducing
\begin{subequations}\label{rotta}
\BEA \label{def3} 
\~I_{ij}&\=& I_{ij}+ \e_{ij}\sp{nl\,,2}\,, \\   \label{isotr} 
\~I_{ij}&\=& \~{\gamma} _{ij}\(3\,W_{ij}-\delta_{ij}W\)\,, \\ 
\label{est5} 
 \~{\gamma} _{ij}(y) &\=& \~{ b }_{ij} \frac{\sqrt{W(y)}}{y }\,,\quad  
\~ b_{ij}=\overline{ b }_{ij}- \frac b 3  \, \delta_{ij}\ .
\EEA\end{subequations}
Recall that tensor $\~I_{ij}$ must have zero trace for any values of $W_{ij}$. This is possible only if $\~b_{xx}= \~ b_{yy}=\tilde  b_{zz}\=\~b \sb d\,, \quad \~b_{xx}\=\~b $, and consequently 
\BE \label{def4}
\~\g_{xx}= \~ \g_{yy}=\tilde  \g_{zz}\=\~\g \sb d\,, \quad \~\g_{xy}\=\~\g\ .
\EE 
Thus,  representation~\Ref{isotr} involves only two free parameters $\~b\sb d $ and $\~b$.   

Equations~(\ref{bal-K}) for $W_{ij}$ with  $ij=xx,\  yy,\ zz$ and $xy$ involve 7 constants $a_{ij}$,  $b$,   $\~b\sb d $ and $\~b$.  Our goal is to formulate the simplest possible model, with a minimal number of adjustable constants. The strategy will be now to use experimental and simulational data, coupled with reasonable physical considerations, to reduce the number of parameters to four, each of which being  responsible for a separate fragment of the underlying physics. 

We should stress that we  neglect in \REF{bal-Ka}  the spatial energy transport term $\epsilon_{\rm tr}$, caused by the tripple-velocity correlations, pressure-velocity correlations and by the viscosity~\cite{00Pope}.   In the high Re$_\lambda$  limit    the density of  turbulent kinetic energy becomes space independent in the log-law region. Accordingly, the spatial transport term  is very small in that log-law region. More detailed analysis, see, e.g. Fig.~3 in Ref.~\cite{DNS}, shows that even for a relatively small Re$_\lambda$ in the log-law turbulent region this term is small with respect to the energy transfer term from scale to scale which is represented by $\gamma  W_{ij}$  in the equations above. On the other hand,  in the viscous sub-layer the mean velocity is  determined by the viscous term and thus the influence of the spatial energy transfer term can be again neglected. To keep the model simple we will neglect $\epsilon_{\rm tr}$ term also in the buffer layer where it is of the same order as the other terms of the model.  The reason for this simplification, which evidently will cause some trouble in the buffer layer, is that the energy balance equations used below become local in space. This is a great advantage of the model, allowing us to advance analytically to obtain a very transparent phenomenology of wall-bounded turbulence. It was already  demonstrated in Ref.~\cite{04LPT} that the simple description ~(\ref{int-sum}) gives a uniformly reasonable description of the rate of the energy dissipation in the entire boundary layer. Here we improve this description further,   effectively accounting for the energy transfer term in the balance equation by an appropriate decrease in the viscous layer parameters $a_{ij}$.

\subsection{\label{ss:sum}Summary of the two versions of the model}
For the sake of further analysis we present the model with the final 
notation:
\BSE\label{MM}
\BEA\label{MM-a} -W_{xy}(y)+\nu_0 S(y) &=& P(y)\,, \\
\label{MM-c}
 [\Gamma_{xx}+ 3 \~{\gamma}\sb d ] W_{xx} &=& \~{\gamma}\sb d  W -2 S 
W_{xy},\qquad 
\\ \label{MM-d}
 [\Gamma_{yy}+ 3 \~{\gamma}\sb d] W_{yy} &=& \~{\gamma}\sb d W \,, 
\qquad \qquad\\
\label{MM-e}
[\Gamma_{zz} + 3 \~{\gamma}\sb d] W_{zz}  &=& \~{\gamma}\sb d W \,,  \\ 
\label{MM-f}
[\Gamma_{xy}+3 \~{\gamma}] W_{xy} &=& - S W_{yy}\,,
\EEA
\ESE
In the traditional theory of wall-bounded turbulence one employs the  
``wall units"  $ u_\tau, \tau$ and
$\ell_\tau$ for the velocity,  time and  length~\cite{00Pope} which for 
a fluid of density $\rho$ are:
\BE\label{wall-u}
    u_\tau\equiv\sqrt{\frac{{P}_0}{\rho}}\,,
     \quad \tau\equiv \frac{\nu_0}
    {P_0}
    \,, \quad \ell_\tau\equiv
    \frac{\nu_0}{\sqrt{ \rho\, {P}_0}}\ .
 \EE%
Using these scales one defines the wall-normalized dimensionless objects
\BEA \nn
y^+&\equiv&\frac{y}{\ell_\tau}\,,\quad
V^+(y)\equiv \frac{V_x(y) }{u_\tau}\,,\quad v^+\Sb T(y)\= \frac{ v\Sb
T(y)}{u_\tau}\,, \quad
etc.\,,\\  \label{dim-l}  
S^+&\equiv& S\,\tau\,, \quad W_{ij}^+(y)\= \frac{W_{ij}(y)}{u_\tau^2}\,,
  \quad  etc.
\EEA %
In our  model we can use the property of locality in space to introduce  ``local units":
\BE\label{loc-u}
    \~ u_\tau(y)\equiv\sqrt{\frac{{P}(y)}{\rho}}\,,
     \ \~ \tau(y)\equiv \frac{\nu_0}
    { P (y)}
    \,, \  \~  \ell_\tau(y)\equiv
    \frac{\nu_0}{\sqrt{ \rho\, {P}(y)}}\,,
 \EE%
similar to traditional wall units~\REF{wall-u}, but with the replacement $P_0\to P(y)$, and ``locally normalized" dimensionless objects, analogous to \REF{dim-l}:
\BEA \label{dim-2}
y^\ddag&\equiv& \frac{y}{ \~\ell_\tau(y)}\,,\quad
v^\ddag\Sb T(y^\ddag) \=  \frac{ v \Sb T(y)}{ \~u_\tau(y)}\,, \br  
S^\ddag(y^\ddag) &\equiv & S\, \~\tau(y)\,, \quad W_{ij}^\ddag(y^\ddag)
\= \frac{W_{ij}(y)}{ \~u_\tau^2(y)}\,,
  \quad  etc.\ .
\EEA %
Then the dimensionless version of \REF{MM-a} reduces to
\BSE \label{1MM}
\BEA \label{1MM-a} - W^\ddag_{xy} + S^\ddag &=& 1\,,\\
\label{1MM-c}
  \( \Gamma_{xx}^\ddag +3 \,\~{\gamma}\sb d ^\ddag\) W_{xx}^\ddag  &=& 
\~{\gamma}\sb d^\ddag \Wd -2 S^\ddag W^\ddag_{xy},\qquad 
\\ \label{1MM-d}
  \(  \Gamma_{yy}^\ddag +3\, \~{\gamma}\sb d\) W^\ddag_{yy} &=&  
\~{\gamma}\sb d\Wd \,, \qquad \qquad\\
\label{1MM-e}
  \( \Gamma_{zz}^\ddag  +3\, \~{\gamma}^\ddag\sb d\) W_{zz} &=&
  \~{\gamma}^\ddag\sb d\Wd \,,  \\ \label{1MM-f}
 \(  \Gamma_{xy}^\ddag + 3\, \gd \)  W^\ddag_{xy} &=&   - S^\ddag 
W^\ddag_{yy}\ .
\EEA
\ESE
Introducing $\vd\= \sqrt {\Wd}$ we can write:
\BSE\label{G}
\BEA\label{Ga}
 \Gamma_{ij}^\ddag&=&  \big( \frac{a_{ij}}{y^\ddag}\big)^2 + 
\frac{b\vd} {y^\ddag}\, \d_{ij}\,, \quad \mbox{for the sum model}\,, \\
\label{Gb}
\Gamma_{ij}^\ddag&=& \sqrt{  \frac{a_{ij}^4}{(y^\ddag)^4 }  
+\frac{b ^2{\vd}^2} {(y^\ddag)^2}\d _{ij}}\,,  \  \mbox{for the root model}\,,\qquad \\ 
\label{Gc}
\~{\gamma}^\ddag\sb d &=&\frac{\~{b}\sb d \vd} {y^\ddag} \,,  \quad
 \~{\gamma}^\ddag  = \frac{\~{b} \vd} {y^\ddag} \,, 
\quad ~\mbox{for  both versions} \ .
 \EEA
 \ESE
\section{\label{s:free}Analysis of the model}
\subsection{\label{ss:gen-an} Solution  of  the 7-parameters   version of the model}
\subsubsection{\label{sss:gen-an}Solutions  in the viscous sub-layer}

The four Eqs.~\Ref{MM-c} -- \Ref{MM-f} can be considered as a homogeneous
``linear" set of equations for $W_{xx}$, $W_{yy}$, $W_{zz}$ and
$W_{xy}$ (with coefficients that are functions of $W$). They can have a
trivial solution $W=0$ for which \REF{MM-a} gives
\BE\label{lam}
S=P_0/\nu_0\,,\quad V=y P_0/\nu_0 \quad W_{ij}=0\,,\ \Rightarrow\
 \mbox{Laminal layer}. 
\EE
The complete absence of turbulent activity in the viscous layer in our
model is a consequence of leaving out the energy transport in physical 
space.

\subsubsection{\label{sss:gen-an} Analysis of turbulent solution}
Equations (\ref{1MM-c} -- \ref{1MM-f}) have
non-trivial ``turbulent solution"  with $\Wd\ne 0$:
\BSE\label{sol3}
\BEA \label{sol3a}
\Wd_{xx}&=&\frac{\Wd}2 \Big[\frac{\Gd_{yy}+\gd\sb d}{\Gd_{yy}
+3\,\gd\sb d}+\frac{\Gd_{zz}+\gd\sb d}{\Gd_{zz}
+3\,\gd\sb d}\Big]\,, \\ \label{sol3b}
\Wd_{yy}&=& \frac{\Wd \,\gd\sb d}{\Gd_{yy}
+3\,\gd\sb d}\,, \quad  \Wd_{zz} =  \frac{\Wd \, \gd\sb d}{\Gd_{zz}
+3\,\gd\sb d}\,, \\ \label{sol3c}
\Wd_{xy}&=&\frac{ -\Wd  \Sd\, \gd\sb d }{\big(\Gd_{xy}
+3\, \gd \big) \big(\Gd_{yy}
+3\, \gd\sb d\big)}\,,
\EEA
if its determinant $\D$ vanishes. The solvability condition $\D=0$ gives:
\BEA\label{g-sol1}
 \(\Sd\)^2&=&\frac{\Gd_{xy}+3\, \gd_{xy}}{2 \gd_{yy}\big(\Gd_{zz}+3\, 
\gd_{zz}\big)}
\Big[\Gd_{xx}\Gd_{yy}\Gd_{zz} \\ \nn &&+ 
2\,\gd\sb d \big(\Gd_{xx}\Gd_{yy} +\Gd_{xx} \Gd_{zz} + 
 \Gd_{yy}\Gd_{zz}
\big )
 \\ \nn &&+  3\,(\gd \sb d)^2   \big(\Gd_{xx} +  \Gd_{yy} 
+ \Gd_{zz}\big )
\Big]\ .
\EEA
\ESE
Substitution $\Wd_{xy}$ and $\Sd$ in Eq.~(\ref{1MM-a}) gives a closed 
equation for the function $\Wd(\yd)$, (or for $\vd\= \sqrt {\Wd}$). To 
present the resulting Eq. in explicit form, introduce
\BEA\label{def2}
A(\vd) &\=& \Sd \big / \vd\,,\quad B(\vd)\= - \Wd_{xy} \big /  \Sd \,  
\vd\,, \br
R_{ij}&\=& \Gd_{ij}\big / \vd\,,\quad \~{r}\sb d \= \gd\sb d \big / \vd\,,
\quad \~{r} \= \gd  \big / \vd\ .
\EEA
Using Eqs.~(\ref{g-sol1}) and (\ref{sol3c})  we find:
\BEA\label{iter4}
 A^2(\vd)&=&\frac{R_{xy}+3\, \~r }{2 \~r\sb d \big(R_{zz}+3\, 
\~r\sb d\big)}
\Big[R_{xx}R_{yy}R_{zz} \\ \nn &&+ 
2\,\~r \sb d\big(R_{xx}R_{yy} +R_{xx} R_{zz} +  R_{yy}R_{zz}
\big )
\\ \nn &&+  3\,\~r \sb d^2 \big(R_{xx} + R_{yy} 
+ R_{zz}\big )
\Big]\,, \br
B(\vd)&=&\frac{ \~r \sb d }{\big(R_{xy}
+3\, \~r \big) \big(R_{yy}
+3\,  \~r \sb d\big)}\ .
\EEA
Now Eq.~(\ref{1MM-a}) can be presented as 
\BE\label{iter5}
 A (\vd)\vd\big[1+ B (\vd)\vd \big]=1\ .
 \EE

Together with Eqs.~(\ref{sol3}) this provides the full solution of 
Eqs.~(\ref{1MM}).

\subsubsection{\label{sss:turb} Outer layer, $\yd >  50$}
In the outer layer, far away from the wall, all the viscous terms in 
Eqs.~(\ref{1MM}) can be neglected. In this case Eqs.~(\ref{G}) for both 
the sum and the root minimal models give:
\BE\label{asym}
\Gd_{ij}\Rightarrow \gd\d _{ij}\,,
\EE
and Eqs.~(\ref{sol3b}--\ref{sol3b})  and the solution \Ref{turb1}
simplifies drastically:
\BE\label{res1}
\Wd_{yy} =   \Wd_{zz} =  \frac{\Wd \, \~b\sb d }{b 
+3\, \~b\sb d }\ .
\EE
By analyzing results of experiments and numerical simulations (as discussed in  Sec.~\ref{s:exp}) we found that in the outer layer $\Wd_{xx}=\frac 12\Wd$, and $\Wd_{yy}=W_{zz}=\frac14 \Wd$. The model reproduces these findings if we choose:
\BE\label{rel-1}b =\~{b}\sb d  \ .
\EE
Using this relation and solution of (\ref{1MM-a}):  $W^\ddag_{xy}=-1$,  in the rest of 
Eqs.~(\ref{1MM}),  one finds
\BE\label{Rel}
\Wd=  \sqrt{\frac{ 24  \~b }{b}} \,, \quad
\frac 1{\kappa}= \(\frac{b } 2\)^{1/4}
(3\, \~ b )^{3/4}\ .
\EE
Here $\kappa$  is nothing but the von-K\'arm\'an constant, that
determines the slope of the logarithmic mean velocity profile in the
log-law turbulent region:
\begin{eqnarray}\label{K-prof}
 \Vp(y^+) &=&\kappa^{-1}\ln y^+ + C\,,
 \quad{\rm for}~ z^+ \gtrsim 30 \,,\\ \nonumber
\kappa&\approx& 0.436\,, \quad C\approx 6.13 \ .
\end{eqnarray}%
The experimental value of $\kappa$ and the intercept $C$,
were  taken from  \cite{97ZS}. Using the simulations result $W^+ \approx 6.85 $ of Ref.~\cite {DNS} which is reproduced in Fig.~\ref{f:comp-meanV},   we find
\BE\label{conts}
b\approx 0.256\,, \quad \~ b \approx 0.500  \ .
\EE

\subsubsection{\label{ss:4-par}  Reduction  of the number parameters: the 
minimal model} 
The parameters $a_{ij}$ are responsible  for the  difference  between the energy  dissipation and the energy transfer  in the viscous sub-layer.  To further simplify  the model we reduce the number of independent parameters  $a_{ij}$ from 4 ($a_{xx}$, $a_{yy}$, $a_{zz}$ and  $a_{xy}$) to two, denoted as $a$ and $\~a$.  Among various possibilities (including $a_{xx}=a_{zz}=a$, $a_{yy}=\~a$, $a_{xy}=(a+\~ a)/2$)   we choose a parametrization  similar to the situation with the outer layer parameters:
\BE \label{res3}
 a_{ii}=a\,, \qquad a_{xy}=\~a \ .
\EE
The analytical solution given in next Sec.~\ref{ss:turb}  simplifies considerably with the 4-parameter version of the model.   A further simplification $a=\~a$ could be considered, but we rule it out since it  yields a monotonic dependence of the turbulent kinetic energy $W^+(y^+)$ with $y^+$, while experimentally there is a pronounced peak of $W^+(y^+)$ in the buffer sub-layer, see  Sec.~\ref{ss:Kin}. We thus consider the four-parameter model as the ``minimal model" (MM).

Below we  will use mostly  the following set of constants:   
\BSE\label{consts}
\BEA\label{a-ta-sum}
a&=&1.0\,, \qquad \~a=10.67  \,, \quad \mbox{ sum-MM},\\ 
\label{a-ta-root}
a&=&1.0\,, \qquad \~a=12.95 \,, \quad \mbox{root-MM},\\  \label{b-tb}
b&=& 0.256\,, \quad \~ b = 0.500\,, \ ~ \mbox{both MMs}.\qquad \EEA\ESE
This choice is based on the analysis of the simulational and experimental data presented in Sec.~\ref{s:exp}.

Notice that eliminating  $b$ from Eqs.~\Ref{Rel} (valid  for both 
sum- and root models) one gets:
\BE\label{rel2}
\vd=12 \kappa \~ b \approx 6\, \kappa\ .
\EE
With the simulational values $\kappa=0.436 $ and ${\Wd}_\infty=6.85 $ this
relationship is valid with a precision that is  better than 1\%.

\subsection{\label{ss:turb} Analysis of the   Minimal Models}
\subsubsection{\label{sss:4par-an}$a,\~a $-parametrization of the 
general solution}
With the minimal parametrization, given by Eq.~(\ref{res3}), the solution~(\ref{sol3}) takes on a simpler form:
\BSE \label{turb1}\BEA\label{turb1a}
\Wd_{yy}&=&\Wd_{zz}=\frac{\vd}{4\, v_4}\, \Wd \,, \quad
\Wd_{xx}=\frac{v_2}{2\, v_4}\,\Wd   \,, \qquad\quad\\ \label{turb1b}
\Wd_{xy }&=&- \frac{\Wd }2  \sqrt{\frac{b\,\vd v _1}{6\, \~ b\,  v_3 v_4
}}\,, \quad 
\Sd=\frac 1 \yd \sqrt{\frac{6\,
b\,\~ b \, v_1v_3v_4}{ \vd}}\ . \qquad\quad  
\EEA\ESE
Here we introduced the following short-hand notations $v_j$ for the sum-MM:
\BSE\BEA\label{turb1c}
v_1 &\=&\vd+ \frac{  a^2}{b\, \yd}\,, \ \qquad v_2\=\vd+
\frac{  a^2}{2\,  b\, \yd}\,,\br
v_3 &\=&\vd+ \frac{ \~  a^2}{3\,\~b\,    \yd}\,, \quad v_4\=\vd+ \frac{  
a^2}{4 \, b\, \yd}\ .
\EEA
For the root-MM instead of Eq.~(\ref{turb1c})   we take:
\BEA\nn 
v_1 &\=&\sqrt{{\vd}^2+  \frac{  a^4}{\(b\, \yd\)^2} }\,,   \quad 
v_2\=\frac{v_1+
 \vd}2  \,,\  v_4\=\frac{v_1 +  3\, \vd}4 \,, \\ \label{turb1d}
v_3 &\=& \frac{\vd \, b}{\~b}+\sqrt{(\~b-b)^2 {\vd}^2+\frac{\~a^4}{(3\, 
\~b \yd)^2}} \ .\qquad
\EEA\ESE

With the minimal parametrization Eq.~(\ref{iter5}) takes a very 
simple explicit form:%
 \BSE \BE\label{turb1d}
 {\vd}^2 +\frac { 12\,\~b\,  \vd }{\yd } \, r_3
r_4 = \sqrt{\frac{24 \, \~ b\,  r_3 r _4 }{b\,  r_1}}\,,\quad r_j\= 
v_j/\vd\ .
\EE
This form of equation for $\vt$ serves below as a starting point for an 
approximate (iterative) analytical solution. One can also seek an exact 
solution by numerical methods; to this aim it is better to use the 
following form of the same equation:
\BEA\nn
F(\vd,\yd )&\=&\frac b {24\, \~b } \, {\vd}^6 v_1 +{\vd} v_3
v_4 \Big[ \frac { b\,  } y  {\vd}^2 v_1  -1\Big] \\ 
 &&
+ \frac {6\,   b\, \~ b }{y^2}\,  v_1  v_3^2  v_4^2 =0\ . \label{poly}
\EEA\ESE 

Equation~\Ref{poly} has seven roots for the sum-MM, (and 27 for the 
root-MM) but only two of them, denoted as
$\vd_\pm$ are real and positive for large enough $\yd$. These two roots
approach each other upon decreasing the distance from the wall. At
some value of $\yd$ these roots merge:
\BE \label{crit}  \vd _+(y\sb{vs})=  \vd _- (y\sb{vs})\= v_*<
{\vd}_\infty\ .
\EE
The values $y\sb{vs}$ and $v_*$ as functions of the problem parameters
follow from the polynomial (\ref{poly}):
\BE\label{crit1}
F(\vd,\yd)=0\,,\quad \frac{\p F(\vd,\yd)}{\p \vd}=0\ .
\EE
For $\yd<y\sb{vs}$ there are no physical (positive definite) solutions of \REF{poly}. This is a laminar region that was discussed before as the viscous sub-layer. In Tab.~\ref{t:1} we present the corresponding values of $y\sb {vs}$ and $v_*$ for $b=0.256\,, \ \~b=0.5$ and various pairs of $a\,, \  \~a$.   

 \subsubsection{\label{sss:iter}Iterative solution  of  \REF{turb1d} 
for  rms turbulent velocity $\vd(\yd)$ }

To develop further analytic insight we employ an iterative 
procedure to find an approximate solution for $\vd_+(y)$ for all  $\yd > 
y \sb{vs}$. For this goal we forget  for a moment that $r_j$ depends on 
$\vd$,  and consider \REF{turb1d} as a quadratic equation with a 
positive solution:
\BE\label{iter2}
\vd=\sqrt{\sqrt{\frac{24 \,\~ b  \, r_3r_4}{b\, r_1}}+\(\frac {6 \, \~
b\, r_3r_4} \yd \)^2}-\frac {6\, \~b\,  r_3r_4} \yd \ .
\EE
However  $r_j$ does depend on $\vd$. For example, for the sum-MM:
\BEA
r_1(\vd) &=&1 + \frac{a^2}{b\, \yd\vd  \Sb T }\,, \quad
r_2(\vd)=1 + \frac{ a^2}{2\,  b\, \yd \vd }\,,\qquad \br
r_3 (\vd)&\=&1+ \frac{\~ a^2}{3\,\~b\, \yd\vd}\,, \quad r_4(\vd)=1+
\frac{a^2}{4 \, b\,\yd  \vd}\ .
\EEA
Nevertheless, for  very large $\yd$ all $r_j\to 1$ and an asymptotic
solution of ~\Ref{iter2} reproduces the asymptotic value of $\vd=
\vd_\infty=(24\, \~b /b)^{1/4}$, given by \REF{Rel}.

A much better approximation for $\vd(\yd)$   (denoted as $\vd_1$)
is obtained using in \REF{iter2} a $\vd$-independent
 $ r_{j,0}\= r_j(\vd_\infty)$ instead of $r_j=1$:
\BSE\label{iter3}\BE\label{iter3a}
\vd_{1}=\sqrt{\sqrt{\frac{24\, \~b  \, r_{3,0} r_{4,0}}{b\,
r_{1,0}}}+\(\frac {6\, \~ b r_{3,0}r_{4,0}} \yd  \)^2}-\frac {6\,\~b  \,
r_{3,0}r_{4,0}} \yd \ .
\EE
Clearly, this iterative procedure can be prolonged further and one can 
find the velocity $\vd$  at the $n+1$ iteration step, $\vd_{n+1}(\yd)$, using the relations  $r_{j,n}\=
r_j(\vd_{n})$, found with the velocity  $\vd_{n}$
of the  previous, $n$-th step:
\BE\label{iter3b}
\vd_{n+1}=\sqrt{\sqrt{\frac{24\, \~b  \, r_{3,n} r_{4,n}}{b\,
r_{1,n}}}+\(\frac {6\, \~b \,  r_{3,n}r_{4,n}} \yd  \)^2}-\frac {6\,
\~b\,  r_{3,n}r_{4,n}} \yd \ .
\EE\ESE
The numerical verification of the iteration procedure is given in Appendix~\ref{a:iter}. The conclusion is that already the first few iterations are sufficiently accurate for all practical purposes:  often one can use the first iteration and occasionally the second one.

Remarkably,  the first iteration can be formulated directly in terms of
the basic Eqs.~\Ref{1MM} by replacing the turbulent velocity profile
$\vd(\yd)$  in Eqs.~\Ref{G} for $\Gd_{ij}$ and
$\gd _{ij}$ by its asymptotic value in log-law region
$\vd_\infty=\(24\,  \~ b/b \)^{1/4}$.

\subsubsection{\label{sss:main} Iterative solution for the mean velocity 
and Reynolds tensor profiles}

Consider first the resulting plots for the mean velocity profile, 
$V^\ddag_n(\yd)$, computed with the help of turbulent velocity 
$\vd_{n}$  at the $n$-th iteration step:
\BE\label{Vn}
V^\ddag_n(\yd)= y\sb{vs}+\int_{y\sb{vs}}^{\yd} S^\ddag_n(\xi) \, d\,
\xi\,, \quad \yd>y\sb{vs}\ .
\EE
Here $ S^\ddag_n(x)$ denotes $ S^\ddag(x)$, given by Eqs.~\Ref{turb1},
with $\vd=\vd_{n}$. Figure~\ref{f:Vn} displays plots of
$V^\ddag_n(\yd)$ for $n=1,\,2,\,3,\,4 \,$ and the ``exact"
(numerical) result $V^\ddag(\yd)$. All the plots almost coincide within
the linewidth. This means that for the purpose of computing $V^\ddag(\yd)$ one can use
the first approximation $\vd_{1}$ given by 
\REF{iter3a} instead of  the exact solution $\vd$.

\begin{figure}
 \centering\includegraphics[width=0.48 \textwidth]{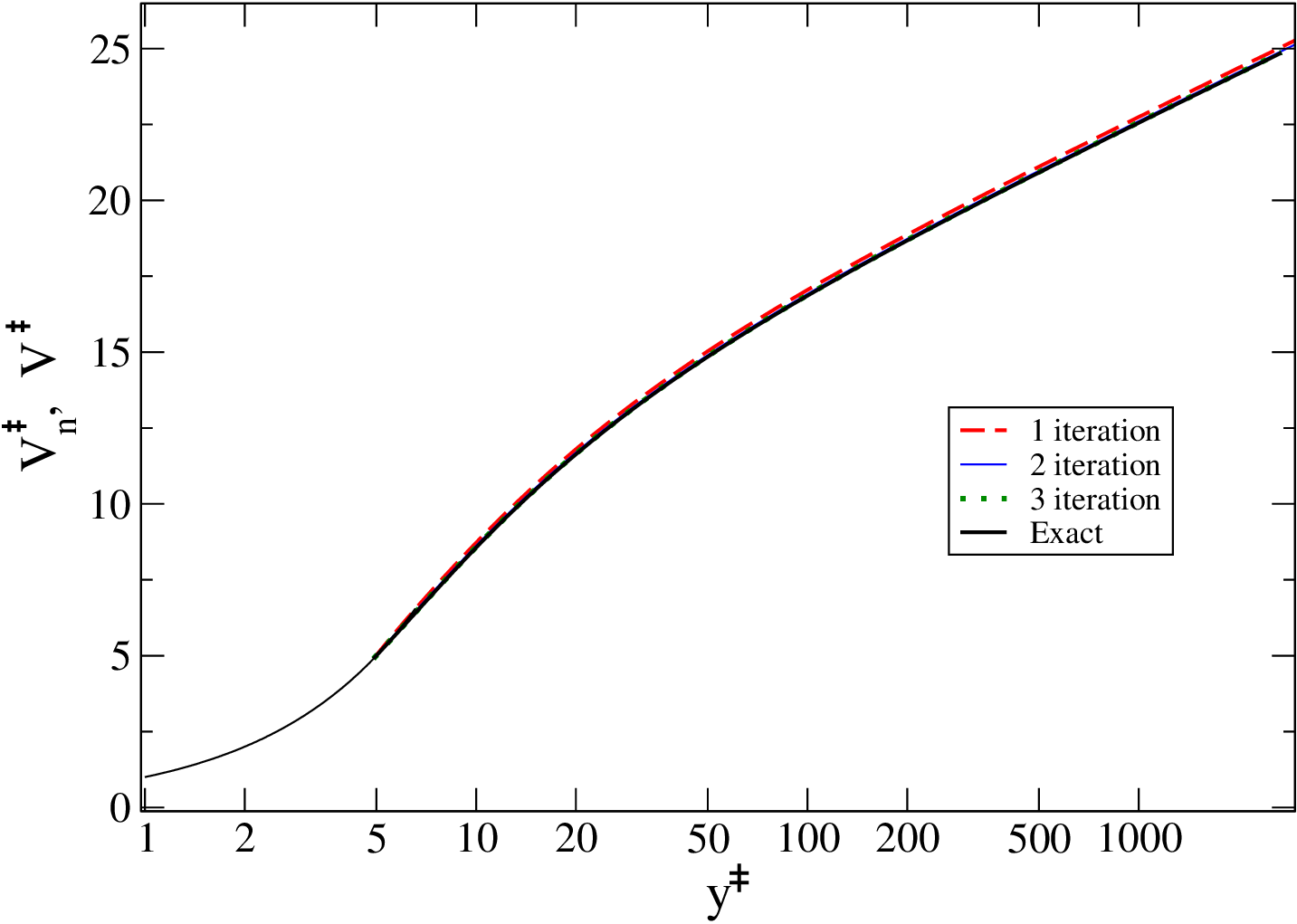} 
 \caption{\label{f:Vn}
  Log-plots of the exact solution for mean velocity profile 
$V^\ddag(\yd)$  and approximate profiles  $V_n^\ddag(\yd)$, computed on 
the $n$-th iteration step for $n=1,2,3,4$ for sum-MM with constants, 
taken from \REF{conts}.  All plots practically coincides within the 
linewidth.
 } \end{figure}

Next we present in Fig.~\ref{f:RnRx} log-plots for  the trace of the
Reynolds-stress tensor $\Wd_{n}(\yd)$,  (computed with the $n$-th iteration
step for $n=1,2,3,4$) together with the ``exact" numerical solution 
$\Wd(\yd)$ for the sum-MM.     Evidently, the iterative 
procedure for the kinetic energy does not converge as rapidly as for the mean velocity profile: one can distinguish the plots of  $\Wd_1(\yd)$, 
$\Wd_2(\yd)$ and $\Wd_3(\yd)$;  the plots of $\Wd_3(\yd)$ and  $\Wd_4(\yd)$  coincide within the line width. Nevertheless,  for $\yd>5 $ (i.e. in   
the buffer layer and in the  outer region)  alreadly the first iterative solution 
provides a very reasonable approximation to the exact solution  for 
the kinetic energy profile.}

\begin{figure}
 \centering\includegraphics[width=0.48 \textwidth]{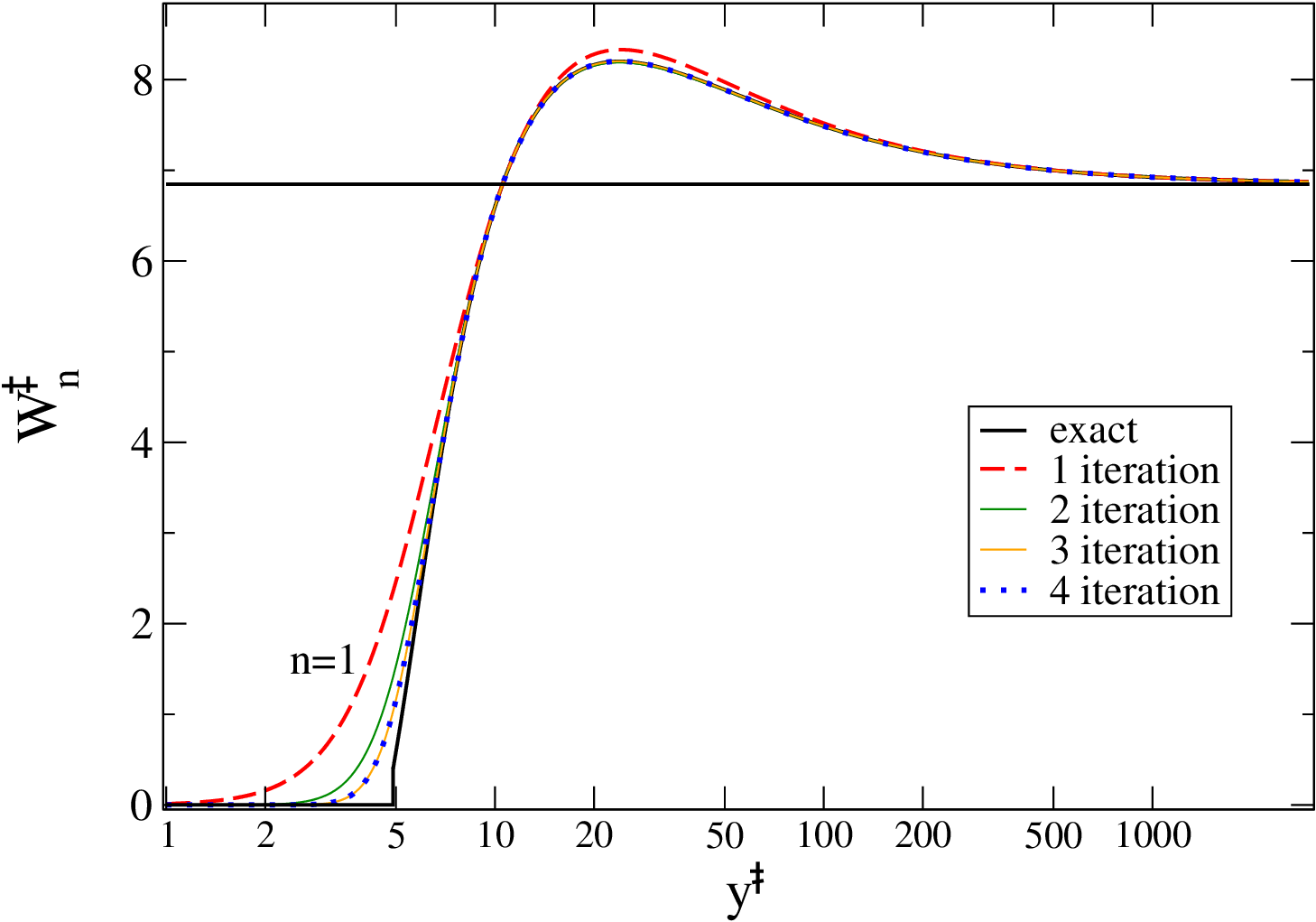}
\caption{\label{f:RnRx} Log-plots of the iterative $\Wd_n$, $n=1\ -\ 4$  and ``exact'' numerical solutions (thick solid line) for 
trace of the Reynolds tensor in  sum-MM with  constants,  taken from 
\REF{conts}.  Plots for $\Wd_3$ and $\Wd_4$ coincide within the line width.
}
\end{figure}
\section{\label{s:exp} Analysis of the numerical and experimental data 
and comparison with the model prediction}
 In this Section we analyze  and compare  the predictions of the minimal models to results   of  experiments and comprehensive direct numerical simulations  of high Re channel flows. We refer to results that were made available in the public domain by  R.~G.~Moser, J.~Kim, and N.~N.~Mansour~\cite{DNS},  to  Large-Eddy-Simulation performed by C. Casciola \cite{LES},  and to recent laboratory experiments in a vertical water tunnel by A. Angrawal, L. Djenidi and R.A. Antonia~\cite{Exp}.   The  choice of the outer layer parameters $b_{ij}$ and $\~b_{ij}$ is based on our analysis of the anisotropy  in the log-low region, presented in  Sec.~\ref{ss:aniz}. The relation between the viscous layer parameters  $a$ vs. $\~a$ is based on the comparison between the DNS and the model mean velocity profiles, presented in Sec.~\ref{ss:meanV}. The final choice of $a$ and $\~a$ is motivated by the DNS data for the kinetic profile which is compared with the model prediction in Sec.~\ref{ss:Kin}.  Section~\ref{ss:Wxy-Wii} is devoted to the comparison of the model results with the DNS profiles of the Reynolds stress $\Wd_{xy}$ and partial kinetic energies $\Wd_{xx}$,   $\Wd_{yy}$, $\Wd_{zz}$.


\subsection{\label{ss:aniz} Anisotropy of the log-layer: Relative partial kinetic 
energies $R_{xx}$,  $R_{yy} $, $R_{zz}$ in the outer layer}

The anisotropy of turbulent boundary layer can be characterized by the 
dimensionless ratios
\BE\label{t-Rii}
R_{ii}(\yp)\= \frac{ W_{ii}(\yp)}{ W(\yp)}=\frac{ \Wp_{ii}(\yp)}{
\Wp(\yp)}\,.
\EE 
This anisotropy plays an important role in various phenomena and was a subject of experimental and theoretical concern for many decades, see, e.g.~\cite{MY,00Pope}.  Nevertheless, up to now the dispersion of results on the subject appears quite large. There is a widely spread opinion, based on atmospheric measurements, that the wall-normal turbulent fluctuations $W_{yy}$ are much smaller than the other ones. For example,   Monin and   Yaglom~\cite{MY}   reported that for a neutrally stratified log-boundary layer $R_{xx}=54\%$, $R_{yy}=6\%$ and $R_{zz}=40\%$.  This  contradicts recent DNS results  for Re$_\lambda$=590 which are available in Ref.~\cite{DNS}, as  shown in Fig.~\ref{f:DNS-RelRii}.   Note that there  is a  region about $100 < \yp < \frac23 \mbox{Re}_\lambda$ where the plots of $R_{ii}(\yp)$  are nearly horizontal, as expected in the log-law region. From these plots we can conclude that is this region $R_{xx}\approx 53\% $ which is close to the value $54\%$, stated in~\cite{MY}. Nevertheless, the DNS data for $R_{yy}$ and $\R_{zz}$ are completely different. From Fig.~\ref{f:DNS-RelRii} one gets $\R_{yy}\approx 22\%$ and $\R_{zz}\approx 27\%$. Thus   $\R_{yy}$ can be considered roughly equal to  $\R_{zz}$.   We should mention here that various models of turbulent boundary layers give $\R_{yy}=\R_{zz}$ in the asymptotic log-law region.  We propose that the difference between $\R_{yy}$ and $\R_{zz}$ which is observed in Fig.~\ref{f:DNS-RelRii}  is due to the effect of the energy transfer. This effect  practically  vanishes in the asymptotic limit Re$_\lambda\to \infty$, but is still present at values of Re$_\lambda$ which are available in DNS~\cite{DNS}. Indeed, for both values of Re$_\lambda$ shown in Fig.~\ref{f:DNS-RelRii}, $W_{yy}=W_{zz}$ in the center of the channel, where the energy flux vanishes by symmetry. Clearly, there is no energy flux also in  space homogeneous cases, for example for a constant shear flow, in which, according to the model, one should expect $W_{yy}=W_{zz}$ in the entire space.

Our expectation that  $W_{yy}=W_{zz}$, which is based on symmetry considerations,  is confirmed by the Large Eddy Simulation (LES) of the constant shear flow~\cite{LES}. As one sees in   Fig.~\ref{f:LES} in this flow $\R_{xx}\approx 0.46$, while $W_{yy}\approx W_{zz}\approx 0.27$.

 As stated,  for sufficiently large values of Re$_\lambda$ the energy transfer terms should almost vanish in the log-law region and, according to our model,  one can expect  in that region $W_{yy}=W_{zz}$ also in the channel flow. This viewpoint was confirmed in the aforementioned laboratory experiment ~\cite{Exp} in a vertical water channel with  Re$_\lambda=1000$, reproduced in Fig.~\ref{f:exp}. The experimental  values of $\R_{xx}$, $\R_{yy}$ and  $\R_{zz}$ in the log-law turbulent region are in the excellent  quantitative  agreement with the values $\R_{xx}=0.5$ and $\R_{yy}=\R_{zz}=0.25$ shown in  Figs.~\ref{f:DNS-RelRii} and \ref{f:exp} by horizontal dashed   lines.

Table~\ref{t:2} summaries the DNS, LES and experimental values of the relative  kinetic energies in comparison with the model expectation. The conclusion is that, in contradiction with the old and still wide spread viewpoint~\cite{MY} that the wall-normal turbulent activity is strongly suppressed, $\R_{yy}<0.1$, the turbulent kinetic energy in the log-law region is distributed in a very simple manner: the stream-wise component contains a half of total energy, $\R_{xx}=\frac12$ and the rest is equally distributed between the wall-normal and span-wise components:     $\R_{yy}= \R_{zz}= \frac14$. As shown in Sec.~\ref{ss:4-par}, this very simple energy distribution is predicted by the minimal model if one assumes that the characteristic nonlinear times scales in the energy transfer term and in the return-to-isotropy term  are identical.  

 Thus anisotropy predicted by our minimal model agree reasonably accurately with those obtained from DNS, LES and vertical water channel. However, we must admit that all these  are not yet the nature. Indeed DNS, LES and lab experiments, done at relatively modest Re$_\lambda$ impose limits on the low-frequency intervals in the spectra of the streamwise and the transverse velocity components, because of side walls or periodic boundary conditions. In other words, it must not be ruled out that DNS, as well as lab experiments, cat off the largest-scale ejections, observed in the atmosphere in the form of coherent structures, which pump additional streamwise- and transverse-velocity energies into the log-layer. Be it as it may, we leave a detailed discussion of the above problem and geophysical applications of our theory for further work. At the present stage, following our strategy of "pragmatic, task-dependent simplification", we consider our minimal model as definitely relevant to flows in channels. Its extension to geophysical (atmospheric and oceanic) boundary layers needs further efforts.

\begin{figure}
\includegraphics[width=0.48 \textwidth]{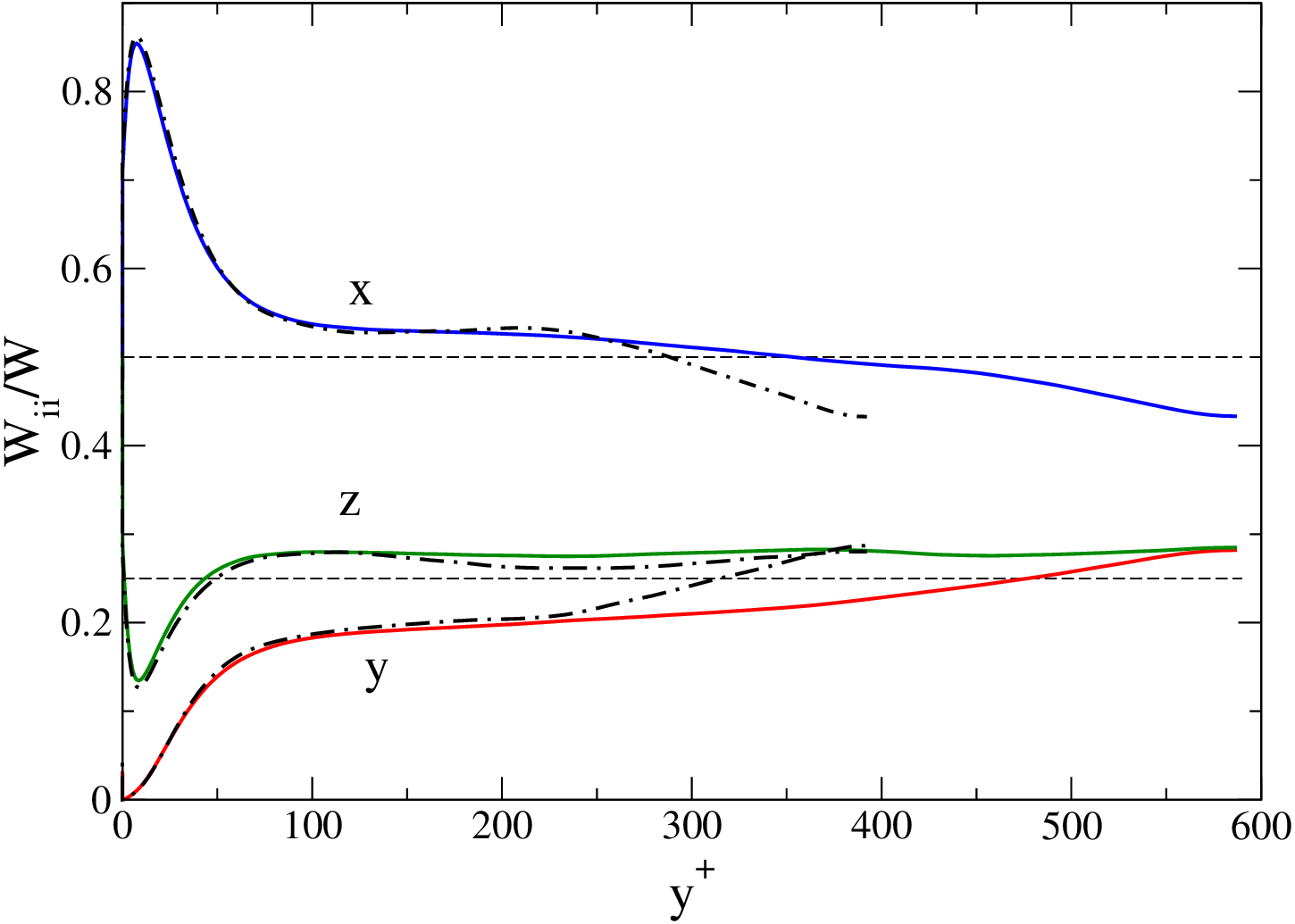}
\caption{\label{f:DNS-RelRii} DNS profiles of the relative kinetic 
energies in the stream-wise, wall-normal and  span-wise directions,  $R_{xx}$,
 $R_{yy}$  and  $R_{zz}$ respectively. Solid lines: Re$_\lambda=590$, 
dot-dashed lines: Re$_\lambda=395$. Horizontal dashed lines show levels 0.5 and 0.25. }
\end{figure}
 
\begin{figure}
 \centering\includegraphics[width=0.48 \textwidth]{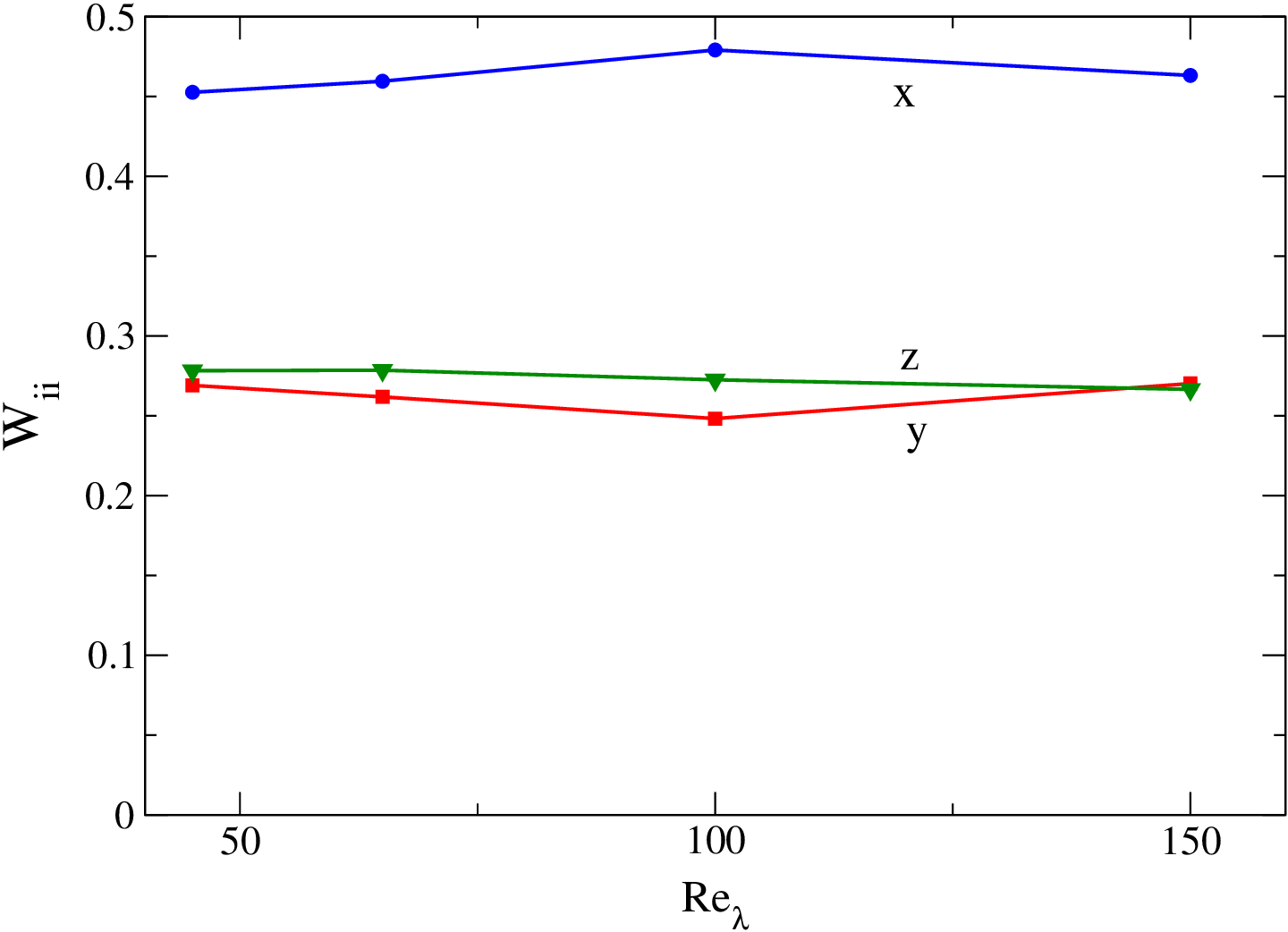}
\caption{\label{f:LES}   Relative components of the Reynolds stress
tensor $\~ W_{ii}$, \REF{t-Rii}, for  for constant-shear flow.
Results of the LES~\cite{LES}. Lines serve only to guide the eye.}
\end{figure}

\begin{figure}
 \centering\includegraphics[width=0.48 \textwidth]{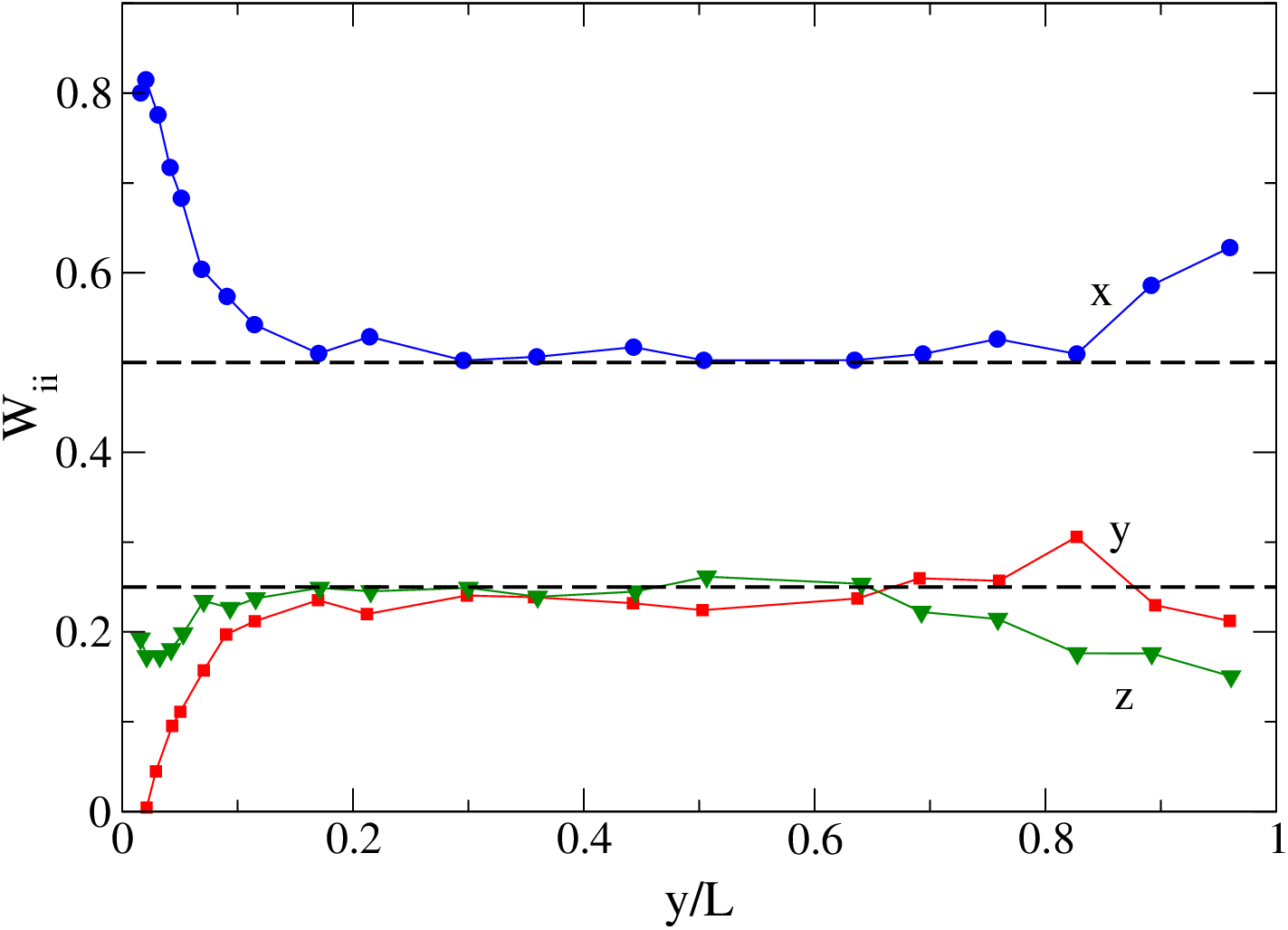}
\caption{\label{f:exp} Experimental profiles of the relative kinetic 
energies $\R_{ii}$ in a vertical water channel with Re$_\lambda=1000$ 
according to Ref.~\cite{Exp}. The solid lines serve to guide the eye.  
The dashed lines show the model prediction in the log-law region 
$\R_{xx}=0.5$ and $\R_{yy}=\R_{zz}=0.25$.}
\end{figure}

\begin{table}
\caption{\label{t:2} Asymptotic values of the relative kinetic 
energies $\R_{ii}$  in the log-law region
$\yd>200$ taken from  DNS (Re$_\lambda=395,\,
590$)~\cite{DNS}, LES \cite{LES} and experiment  in a water
channel with Re$_\lambda=1000$ \cite{Exp}. The last column presents the predictions
of the minimal model.}~\\
\begin{tabular}{||r|c|c|c| c||}
  \hline \hline
  ${R_{ii,\infty}},\ \downarrow ii \downarrow $  & DNS~\cite{DNS}  &
LES~\cite{LES}&  Water channel~\cite{Exp} & Model \\
  \hline
  $xx~$  &  $\approx 0.53$ &$\approx 0.46$ & $0.50\pm 0.01$ & 0.50 \\
$yy~$  & $\approx 0.22$ &$\approx 0.27$  & $ 0.25 \pm 0.02$ &  0.25 \\
 $zz~$ & $\approx 0.27$  &$\approx  0.27$ &  $ 0.25 \pm 0.02$ & 0.25 \\

  \hline\hline
\end{tabular}
\end{table}
\subsection{\label{ss:meanV}Mean velocity profile in  channel flows}

\begin{figure}
 \centering\includegraphics[width=0.48 \textwidth]{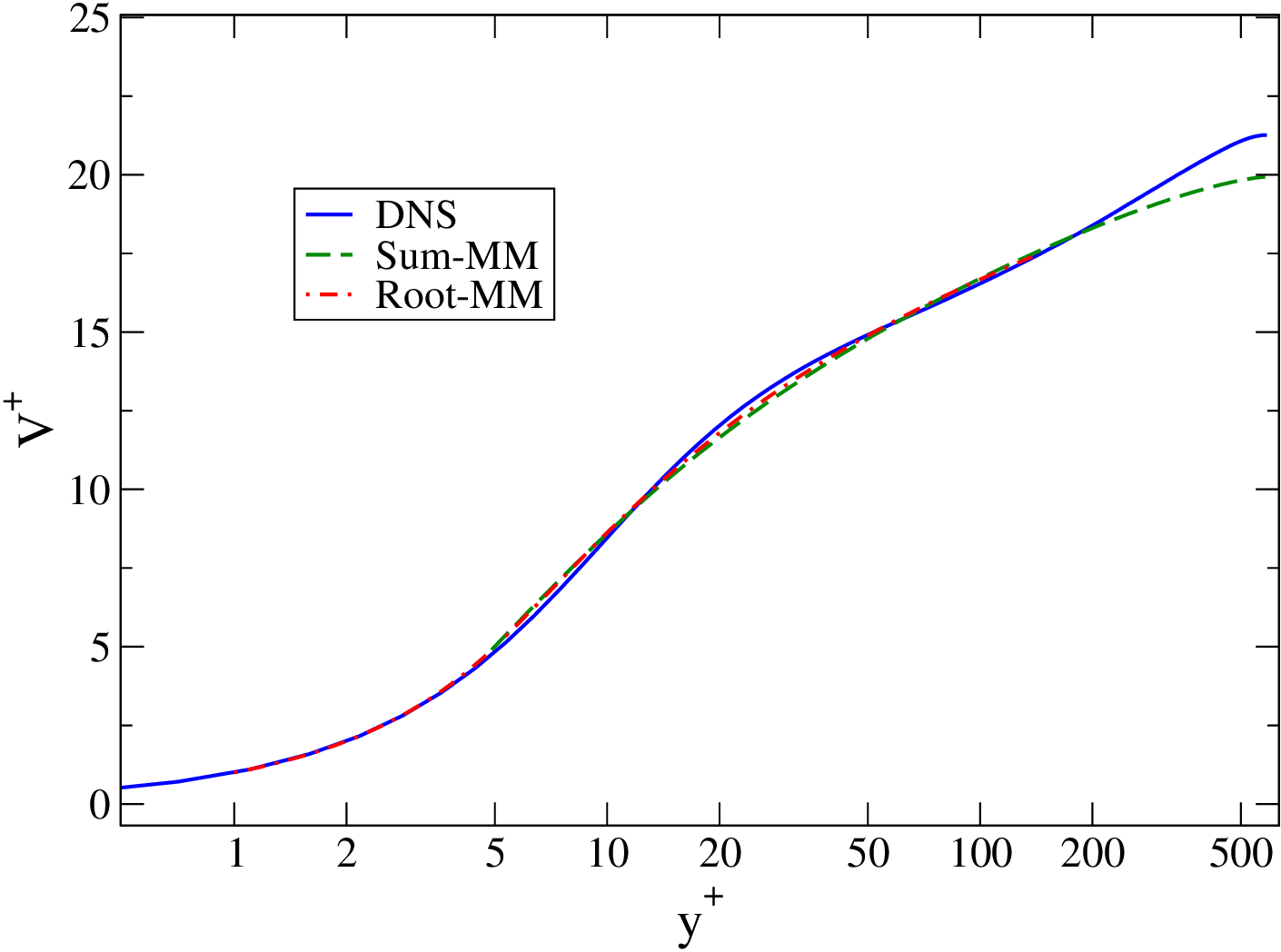}
\caption{\label{f:comp-meanV} Mean velocity
profiles $V^+(y^+)$: The solid blue line -- DNS data~\cite{DNS}
for Re$_\tau=590$, the dashed  green line --  analytical profile
$V^+ (y^+)$ for the sum-MM, and dash-dotted red line for the
root-MM using the parameters ~\REF{conts}. The relative deviation of the
analytical predictions from the DNS data are about $1\%$, smaller
for the root-MM than for the sum-MM.}
\end{figure}

To compute the mean velocity profile  $\Vp(\yp)$ in our
approach we need to connect first $S^+(\yp)$ with $\Sd(\yd)$.
According to the definitions~\Ref{dim-l}~--~\Ref{dim-2}:
\BE\label{rel1}
S^+(\yp)= \(1- \frac{\yp}{\mbox{Re}_\lambda}\)\Sd\(\yp \sqrt{1-
\frac{\yp}{\mbox{Re}_\lambda}}\ \)
\,,
\EE
where 
\BE\label{Re}
\mbox{Re}_\lambda\=L/\ell_\tau\,,
\EE
and $\yp>y\sb{vs}$. For $\yp<\y\sb{vs}$ we can take $S^+(\yp)=1$
and integrate the resulting shear over the distance to the wall
with no-slip boundary condition. The resulting profiles $\Vp(\yp)$
for Re$_\lambda=590$  and the parameters~\Ref{conts}  are  shown in
Fig.~\ref{f:comp-meanV} as a dashed line  for the sum-MM  and as a
dot-dashed line for the root-MM. The DNS profile of~\cite{DNS} for
the same  Re$_\lambda$ is shown as a solid line. There is no
significant difference (less than $1\%$) between these plots in 
the viscous
sublayer, buffer and outer layers, where
$\yp\lesssim 300$ i.e. in about 50\% of the  channel half-width
$L^+=\mbox{Re}_\lambda =590$. This  robustness  of  the mean velocity 
profile $V^+(y^+)$   is a consequence of the fact that $V^+(y^+)$ is an 
integral of the mean shear $S^+$ which is described very 
well both in the viscous and the outer layers.    

Notice that our model does not describe the upward deviation
from the log-low which is observed near the mid-channel (of about 5-6
units in $V^+$, independent on Reynolds number).  We consider this
minor disagreement as an acceptable price for the
simplicity of the minimal model which neglects the energy transport term
toward the centerline of the channel. This transport is the only reason 
for some turbulent activity near the centerline where both
the Reynolds stress $W_{xy}$ and the mean shear $S $ vanish  due to  
symmetry. Just at the center line the source term in our energy equation,  
$ -2 S W_{xy}$, is zero, and the missing energy transport term is felt.

The plots in Fig.~\ref{f:comp-meanV} have a reasonably straight logarithmic region from $\yp \approx 20$ to $\yp \approx 200$. On the other hand, the Reynolds stress profile at the same Re$_\lambda=590$ shown in Fig.~\ref{f:comp-Rx},  has no flat region  at all. Such a flat region is expected in the true asymptotic regime of Re$_\lambda\to\infty$, where $\Wp=-1$.  Therefore if one plots  the model profiles $V^+$ at different Re$_\lambda$ and fits them  by log-linear profiles~\Ref{K-prof} one can get a  Re$_\lambda$-dependence of the ``effective" intercept  in the \vK log-law.  We think that this explains, why measured value of the log-low intercept can depend on the Reynolds number and on the flow geometry (channel vs. pipe): both in DNS and in physical experiments one usually does not reach high enough values of Re$_\lambda$.

\subsection{\label{ss:Kin}  Profiles of the total kinetic energy density 
and the choice of the pair $a,\ \~a$}

The quality of the profiles $\Vp(\yp)$ calls for a bit more thinking. In fact, one find that the minimal model produces practically the same profiles $\Vp(\yp)$ not only for the parameters~\Ref{consts} but for a wide choice of the pairs $a,\ \~a$, for example for $a=2$ and $\~a=8.6$. Actually, for any $0 \leq a \lesssim 4 $  one can find a value of $\~a$ that gives a mean  velocity profile in good agreement with Fig.~\ref{f:comp-meanV}. In other words, in the $(a,\~a)$-plane there exist a long narrow corridor that produces a good quantitative description of $\Vp(\yp)$. Within this corridor there exists a line that provides a  ``best fit" of $\Vp(\yp)$, minimizing  the mean square deviation $\d \Vp$
\BE\label{deviation}
 \delta \Vp\equiv  \sqrt{\<[V\Sb{MM}^+(\yp)-V\Sb{DNS}^+(\yp)]^2\>}
 \EE
of the model prediction  $V\Sb{MM}^+(\yp)$ from the DNS profile $ 
V\Sb{DNS}^+(\yp)$ in the inner region $\yp< 140$. Some of the best pairs 
are given in Table~\ref{t:1} together with  the corresponding  values  of  
$\d \Vp$. Table~\ref{t:1} also presents  values of  $y\sb {vs}$ and 
$v_*$; recall: for $\yd<y\sb {vs}\,,\ \vd=0$, for $\yd=y\sb {vs}+0$ 
there is a jump of $\vd$  from zero to  $\vd=v_*$. The most striking
difference for different $( a,\ \~a)$ pairs  is in the behavior of the 
Reynolds stress profiles $\Wd(\yd)$ that can be used to select the best
values of these parameters.
\begin{table}\caption{\label{t:1} Optimal values of $\~a $ for a given 
value of $a$, that   
$\delta \Vp$ of Eq. (\ref{deviation}). For optimal pairs $(a,\~a)$ in the sum and the root versions 
of the minimal model (denoted as $\sum$ and  $\displaystyle  \sqrt{~~} ~$ 
correspondingly) we present the values $y\sb{vs}$ of $\yd$, 
separating the viscous solution with $\vd=0$ from turbulent regime with 
$v_*\equiv \vd (y\sb{vs}+0) $. The last two columns present the $\yd$ 
position, $y\sb{max}$,  of the maximum of the kinetic energy  and the
corresponding values of $\Wd\sb{max}\equiv \Wd(y\sb{max})$.     }


\begin{tabular}{||c|c|c|c||c|c||c|c||}
  \hline  \hline
MM&   $a$ & $\~a$ & $\delta \Vd$ & $y\sb{vs}$ & $v_*$ & $y\sb{max}$ & 
$\Wd\sb{max}$ \\
  \hline  \hline
 ~$\sum $~&  ~~0.1~~ & ~~10.4~~ & ~~0.29~~ & ~~1.5~~ & ~~0.017~~ & 
~~21~~ & ~~8.33~~ \\  \cline{3-8}
$\sqrt{~~} $& 0.1 & 12.0& 0.25 & 1.1 & 0.008 & 18 & 8.32 \\  \hline
$\sum $ &  0.25 & 10.9 & 0.25 & 2.4 & 0.061 & 22 & 8.48 \\  \cline{3-8}
$\sqrt{~~} $ &  0.25 & 12.1 & 0.25 & 2.1 & 0.016 & 19 & 8.36 \\  \hline
   $\sum $&0.5 & 11.1 & 0.25 & 3.4 & 0.159 & 22 & 8.47 \\ \cline{3-8}
  $\sqrt{~~} $&  0.5 & 12.6 & 0.17 & 3.4 & 0.247 & 19 & 8.50 \\  \hline
   $\sum $&  1.0 & ~10.7 & ~0.22  & 4.8 & 0.401 & ~24.6 & 8.24 \\ 
\cline{3-8}
  $\sqrt{~~} $&  1.0 & ~12.9  & 0.15 & 4.9 & 0.234 & ~19.2 & 8.59 \\  \hline
   $\sum $&  1.5 & ~9.7 & 0.20 & 5.7 & 0.600 & 27 & 7.76 \\  \cline{3-8}
    $\sqrt{~~} $&  1.5 & 12.9 & 0.16 & 4.8 & 0.743 & 19 & 8.58 \\  \hline
   $\sum $&  2.0 & ~8.6 & 0.19 & 6.4 & 0.783 & 27 & 7.26 \\ \cline{3-8}
  $\sqrt{~~} $&  2.0 & 11.8 & 0.17 & 6.7 & 0.743 & 20 & 8.13 \\  \hline
   $\sum $&  4.0 & ~2.9 & 0.21 & 7.2 & 0.630 & $-$ & $-$ \\ \cline{3-8}
   $\sqrt{~~}$&  4.0 & ~6.3& 0.46 & 8.5 & 1.04& $-$ & $-$ \\
  \hline
\end{tabular}\end{table}
\begin{figure}
 \centering\includegraphics[width=0.48 \textwidth]{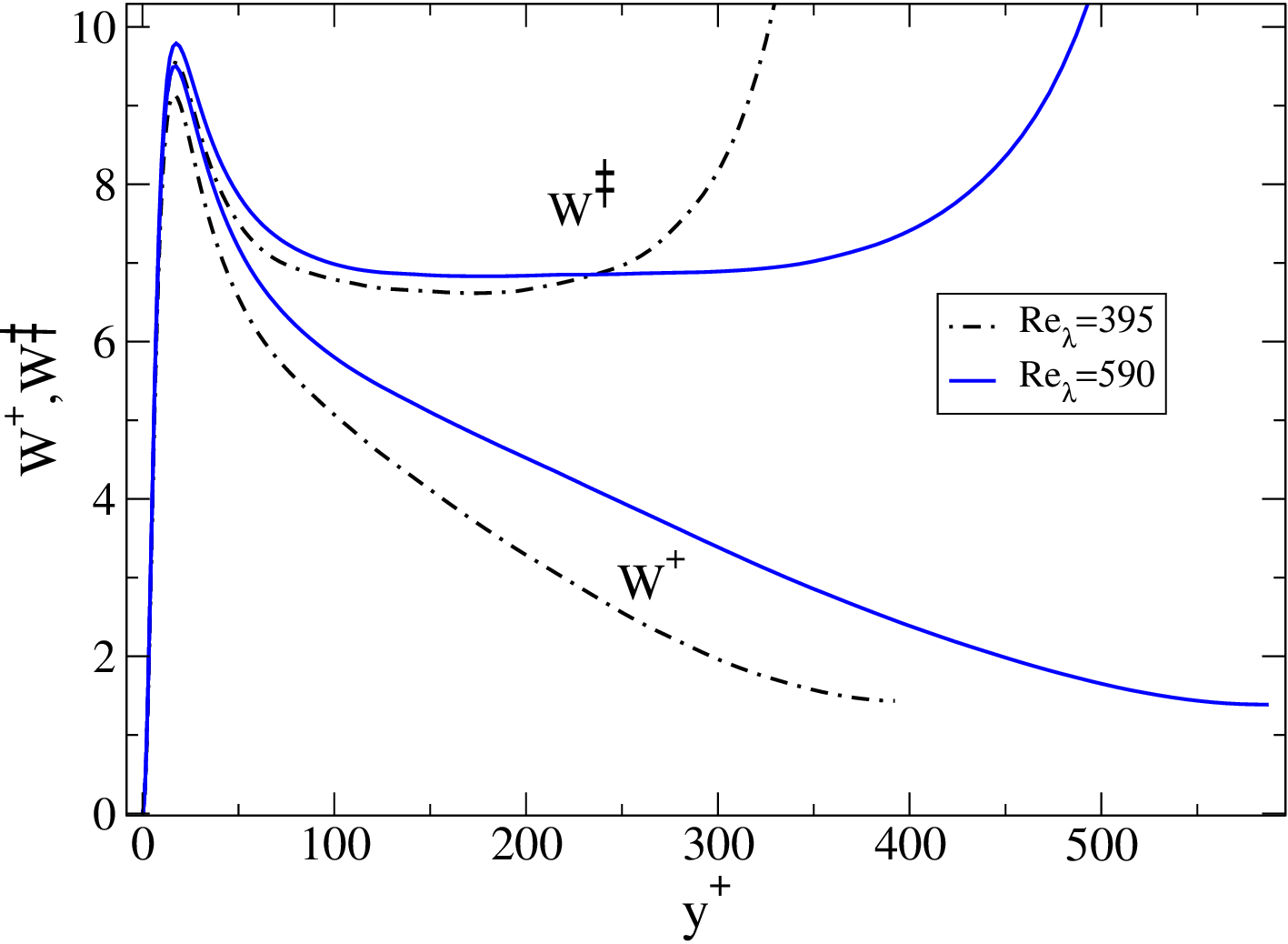}
 \vskip 1cm
 \centering\includegraphics[width=0.48 \textwidth]{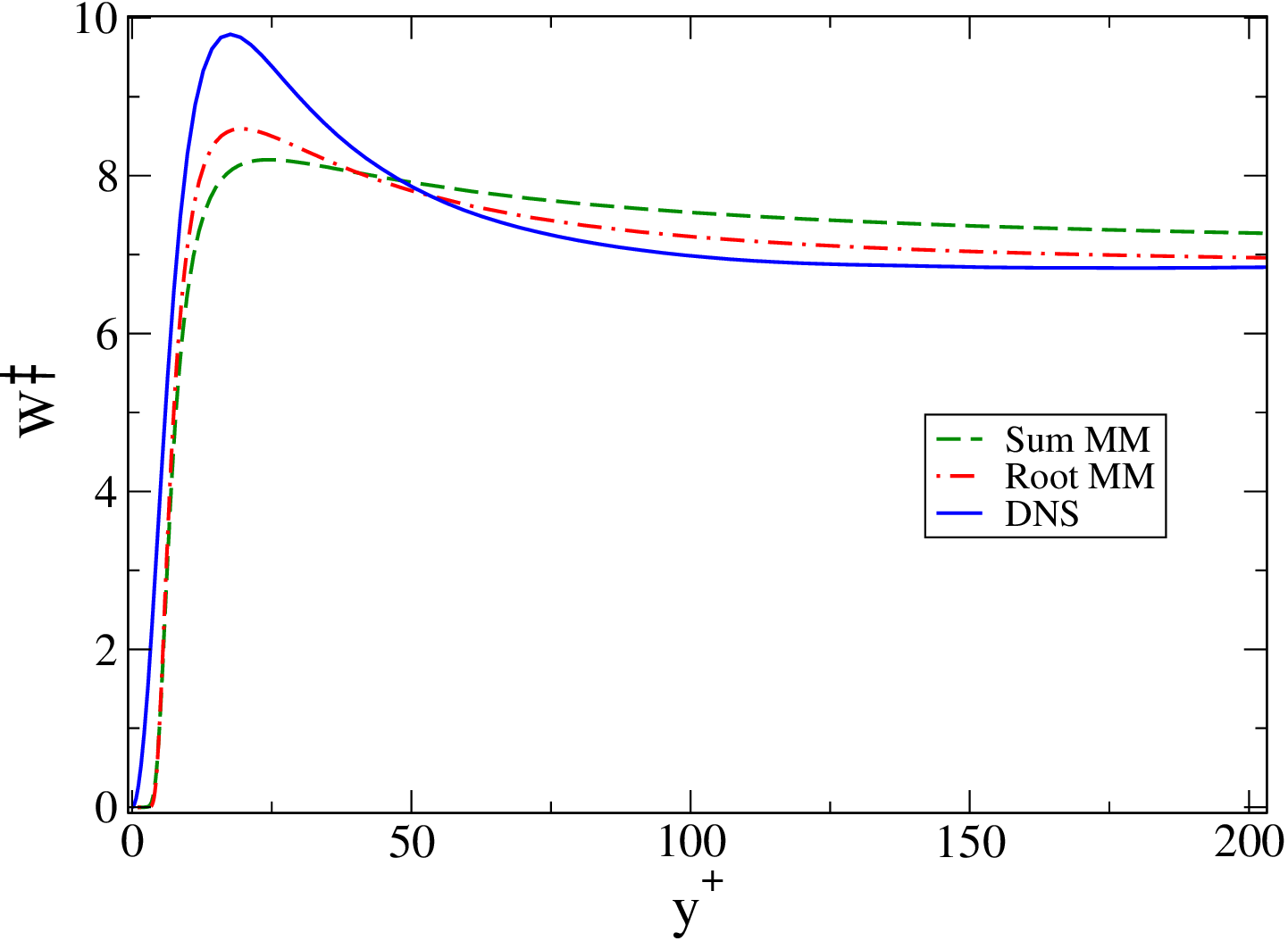}
 \caption{\label{f:4}   Upper panel: DNS profiles of  the trace of the 
Reynolds-stress tensor (twice of the   total kinetic energy densities)  
in two normalization, $\Wp$ and $\Wd$. Solid blue line: Re$_\tau=590$ and 
dot-dashed black line -- for Re$_\tau= 395$.   Lower panel:   Comparison of 
the DNS profile $\Wd(\yp)$ for  Re$_\lambda=590$ (blue solid line) with 
the results of the sum-MM  (green dashed line) and the root-MM (red 
dot-dashed line) with  constants~\REF{consts}. } \end{figure}

Clearly, the minimal model with only 4 fit parameters cannot fit
perfectly the profiles of all the physical quantities that can be
measured. Therefore the actual values of $a$ and $\~a$ should be
determined with a choice of the characteristics of turbulent boundary
layers that we desire to describe best. Foremost in any modeling
should be the mean velocity profile which is of crucial importance in
a wide variety of transport phenomena. Next we opt to fit well the
profile of the kinetic energy density ([or, equivalently, the profile
of the Reynolds stress tensor trace $\Wd(\yd)$] .  Figure~\ref{f:4},
upper panel, shows the DNS profiles of the trace of the Reynold-stress
tensor $\Wp(\yp)$ for Re$_\lambda=590$ (solid lower line) and
Re$_\lambda=395$ (dashed lower line). There are no plateau in these
plots, meaning that these values of Re$_\lambda$ are not large enough
to have a true scale-invariant log-law region. Nevertheless, the plots
of 
\BE 
\Wd(\yp)=(1-\frac{\yp}{{\rm Re}_\lambda})^{-1}\Wp(\yp) 
\EE
(shown in
the same upper panel of Fig.~\ref{f:4}) display clear plateaus,
according to the theoretical prediction for Re$_\lambda\to
\infty$. This means that the decay of $\Wp(\yp)$ is related to the
decrease of the momentum flux $P(y)$ and that the dimensionless ``
$^\ddag$ " variables, \Ref{dim-2}, that use the $y$-dependent value of
the momentum flux $P(y)$ represent the asymptotic physics of the wall
bounded turbulent flow, at lower values of Re$_\lambda$ than the
traditional ``wall units" \Ref{dim-l}, which are based on the wall
value of the momentum flux $P_0$.

To compare the model prediction with simulational results we have to 
relate $\Wd_{ij}(\yd)$  with $\Wp_{ij}(\yp)$ in  channel flows. 
According to Eqs.~\Ref{dim-l}~--~\Ref{dim-2}: 
\BE\label{rel1} 
W_{ij}^+(\yp)= \(1- \frac{\yp}{\mbox{Re}_\lambda}\)\Wd_{ij}\(\yp 
\sqrt{1- \frac{\yp}{\mbox{Re}_\lambda}}\ \) \,, 
\EE
 and similar Eqs. for the its trace  $\Wp(\yp)$. Figure~\ref{f:4} shows a  
peak of $\Wd(\yp)$, $\Wd\sb{max}\equiv 
\Wd(y\sb{max})\approx 9.8$ at $\yp=y\sb{max}\approx 18$.   As one sees 
from the Tabl.~\ref{t:1}, the minimal model reproduces  the peak in  $\Wd(\yd)$ 
with an amplitude of about $8\div 8.6$ for $a \lesssim 2$.  To be
specific  we choose $a=1$ in both versions of the minimal model, sum-MM and root-MM. 
With this choice we plot in Fig.~\ref{f:4}, lower panel,  both 
theoretical profiles,  $\Wd_{_{\sum}}(\yp)$ and 
$\Wd_{_{\sqrt{~}}}(\yp)$, in comparison with the simulational profile  
$\Wd\Sb{DNS}(\yp)$. It appears that the root-MM is in better correspondence 
with the simulation than the sum-MM. However, for the sake of analytic calculations, the 
sum-MM is simpler. Therefore, again, the choice of the version of MM 
depends on what is more important for a particular application: calculational
simplicity or accuracy of fit.

\begin{figure}
 \centering\includegraphics[width=0.48 \textwidth]{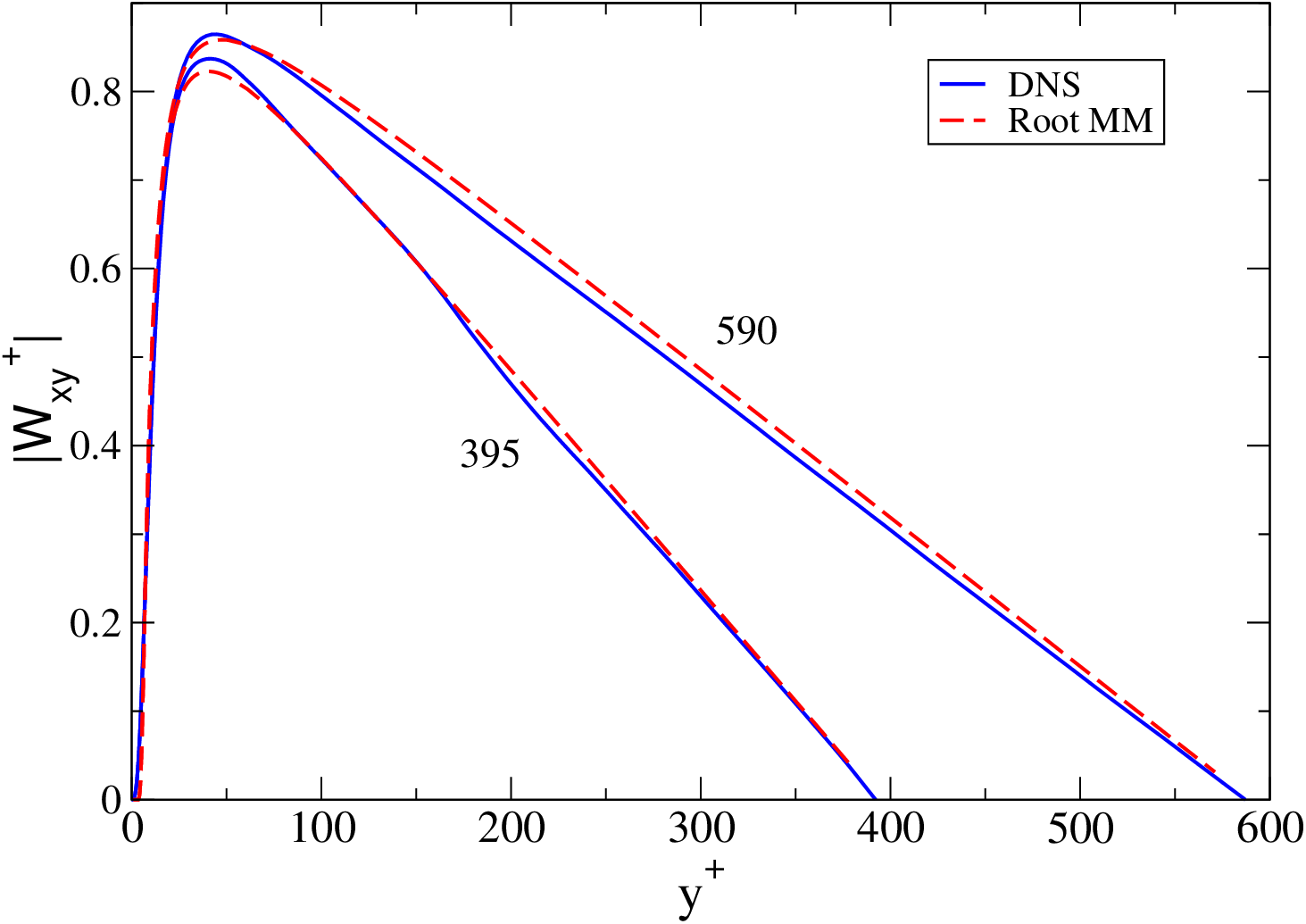}
 \vskip 1.2cm
\centering\includegraphics[width=0.48 \textwidth]{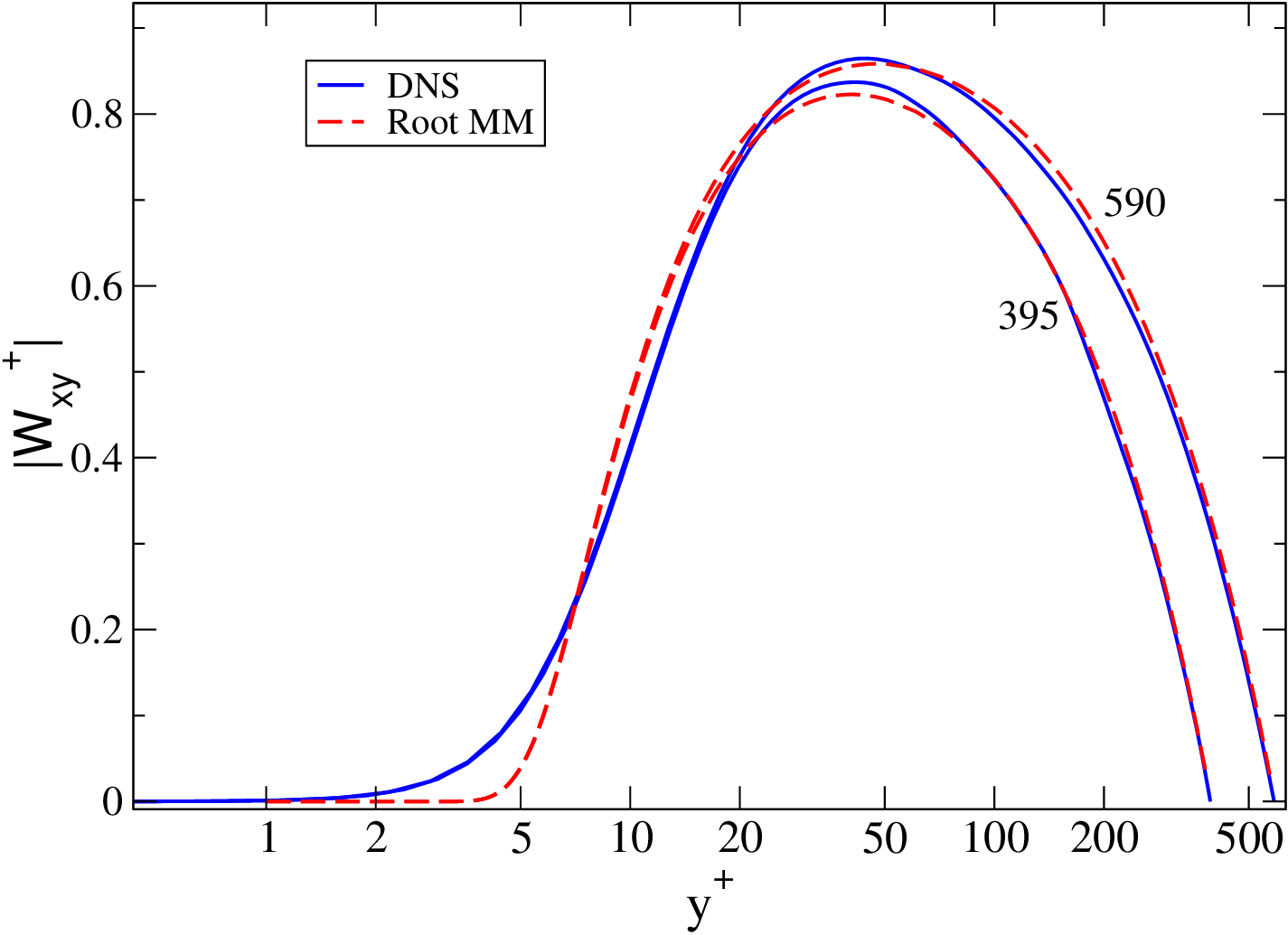}
\caption{\label{f:comp-Rx} Comparison of the DNS Reynolds stress 
profiles $ |\Wp_{xy}|$ (blue solid lines) for  the channel  flow with 
Re$_\lambda=395$ and $590$ with the root-MM  profiles with 
constants~\Ref{consts}), red dashed lines. Upper panel: linear coordinates, 
lower panel: the same plots in the linear-log coordinates. }
\end{figure}

\subsection{\label{ss:Wxy-Wii} Profiles of the Reynolds stress tensor}

In Fig.~\ref{f:comp-Rx}  we present (by solid lines) simulational profiles of 
the Reynolds stress $\Wp_{xy}(\yp)$ for Re$_\lambda=395$ and 
Re$_\lambda=590$ in comparison with the model  predictions (dashed 
lines) for the root-MM. The upper panel shows the comparison in 
linear coordinates, the lower panel in linear-log coordinates,  
stressing the buffer layer region.  In the model profiles we 
used the values of parameters~\Ref{consts},  chosen to fit the simulational
profiles  for  the mean velocity  and the kinetic energy.   In other 
words,  in  comparing the profiles of $\Wp_{xy}(\yp)$ in 
Fig.~\ref{f:comp-Rx} \emph{no further fitting was exercised}.  Having this in mind, we consider the agreement as very 
encouraging.   The only difference between the model predictions and the simulational profiles of 
$ \Wp_{xy}(\yp)$ is  in a steeper front of the model profiles for $\yp<20$. 
This is again because the model does not account for  the energy transfer that 
can only flatten the front. As already mentioned,   even for Re=590 the 
maximum value of  the Reynolds stress does not reach it asymptotical 
value $|\Wp_{xy}|=1$, as it should in the true log-law region.   The 
corresponding comparison for the sum-MM looks very similar and is therefore 
not shown.

\begin{figure}
 \centering\includegraphics[width=0.48 \textwidth]{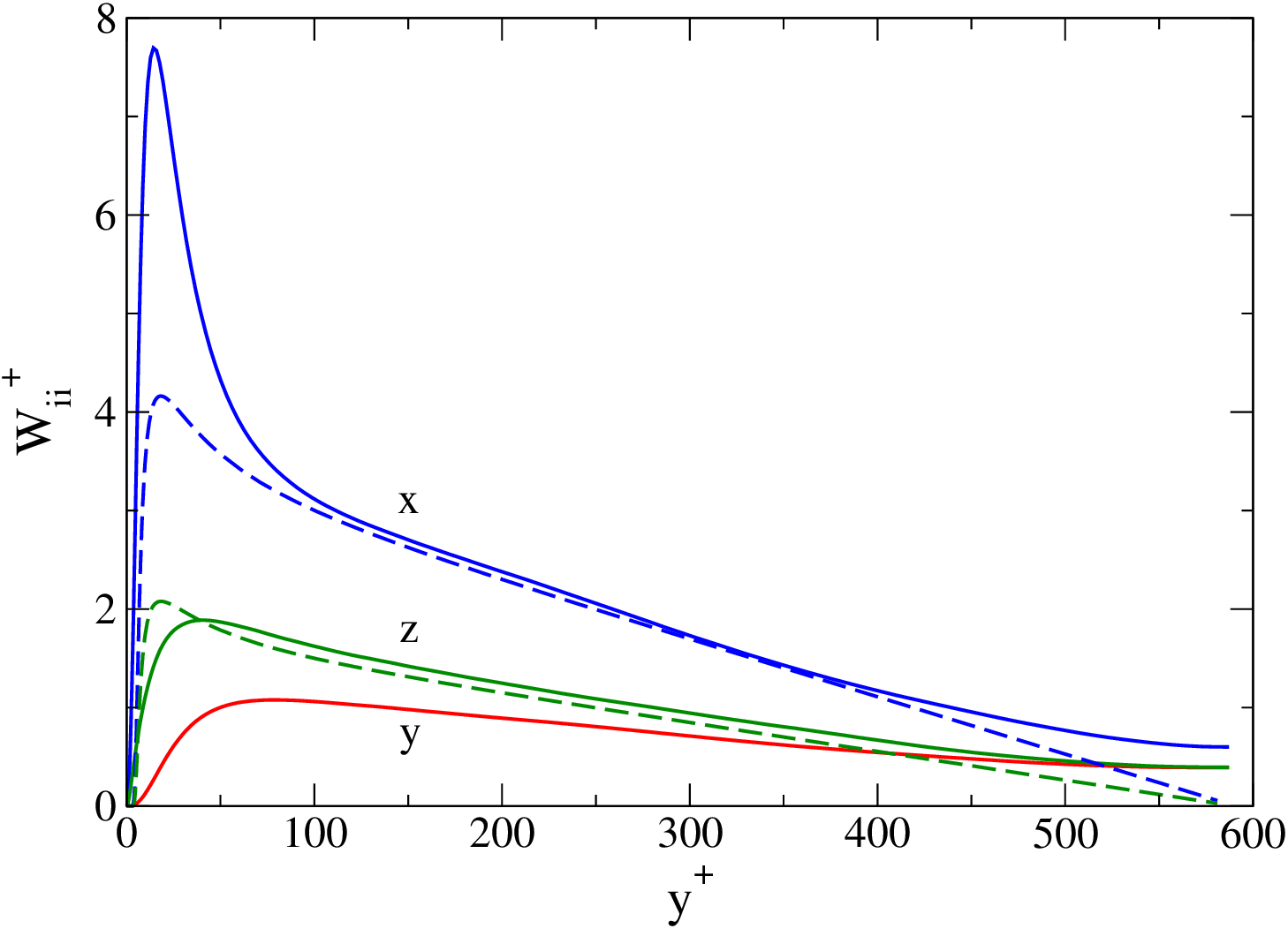}
 \caption{\label{f:comp-Rii} Comparison of the DNS (solid lines) and root-MM results (dashed lines) for  partial kinetic energies profiles in the channel flow with Re$_\lambda=590$. The model parameters are taken from~\Ref{consts}.}
\end{figure} 
\begin{figure}
 \centering\includegraphics[width=0.48 \textwidth]{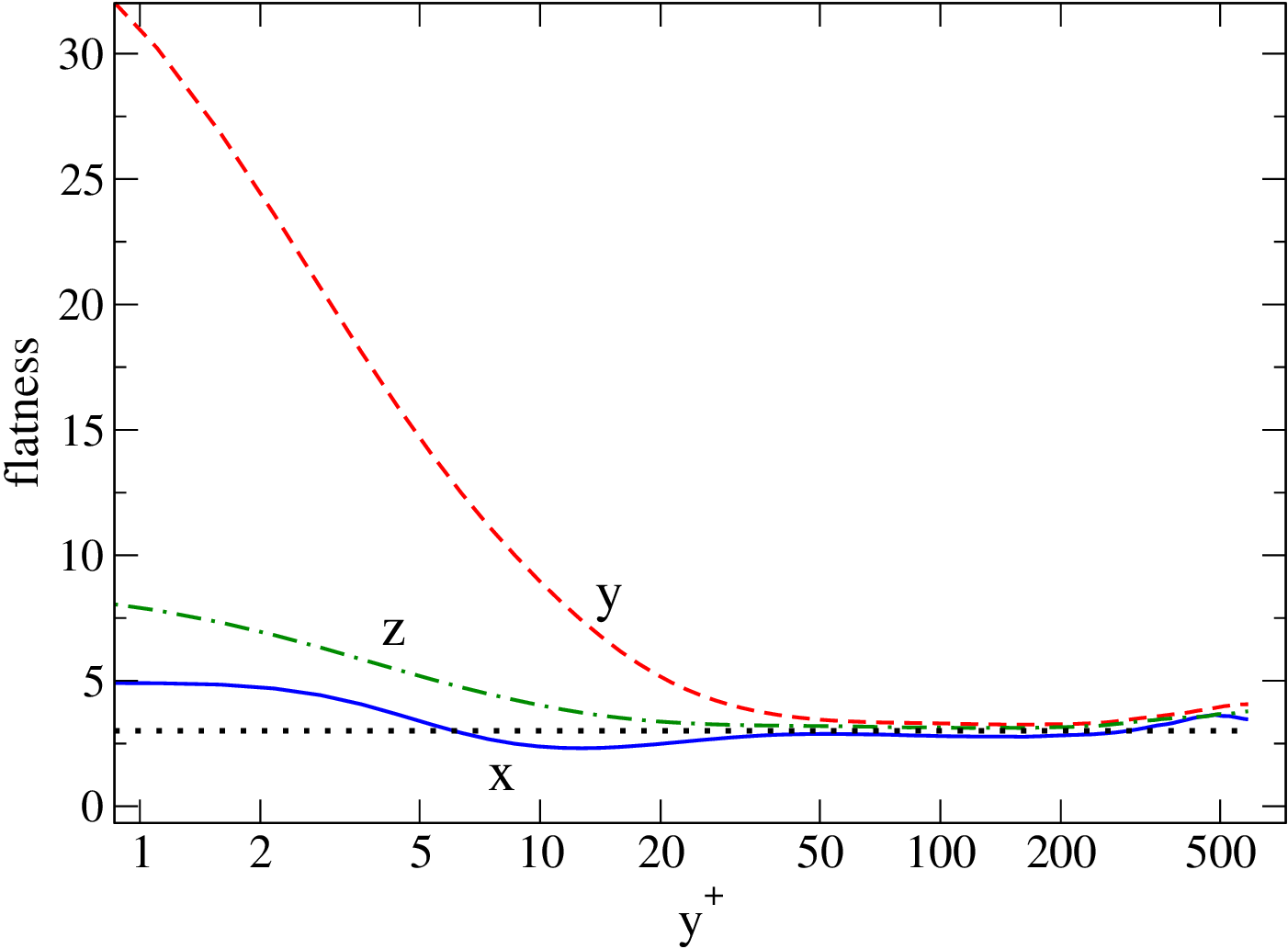}
\caption{\label{f:flatness} Profiles of the flatness of the $x$, $y$ and 
$z$ components of the turbulent velocity fluctuations, DNS 
data~\cite{DNS}  for Re$_\tau=590$. Horizontal dotted line shows Gaussian value  for the flatness, equal to three.}
\end{figure}

Next, we present in Fig.~\ref{f:comp-Rii} the simulational and root-MM
profiles of the diagonal components of the Reynolds-stress tensor,
$\Wp_{ii}(\yp)$ for the channel flow with Re$_\lambda=590$.  Solid
lines present simulational profiles, dashed lines -- the model
profiles.  The stream-wise and span-wise profiles, $\Wp_{xx}(\yp)$ and
$\Wp_{zz}(\yp)$, are in good agreement in the most of the channel,
$70<\yp< 470$, while for the wall-normal component, the model profile
$\Wp_{yy}(\yp)=\Wp_{zz}(\yp)$ and differs from the
simulational one.  The model also predicts semi-quantitatively increase in
the streamwise part of the kinetic energy and the decrease in the
span-wise and wall-normal components in the buffer layer which is
observed in simulations. The physical reason of this is simple: as is
well known, the energy from the mean flow is transferred only to the
stream-wise component of the turbulent fluctuations. Accordingly, in
the model one sees the energy production term ($ -2 S W_{xy}$) only in
the RHS of equation for $W_{xx}$. The energy redistributes between
other components due to ``return-to-isotropy" term $I_{ij}$,
\REF{isotr} with the isotropisation frequency $ \propto 1/ y $. The
relative importance of $I_{ij}$ (in comparison with the energy
relaxation term) decreases toward the wall due to the viscous
contribution $\propto 1/ y^2$. Accordingly, near the wall only a small
part of the kinetic energy can be transferred from the streamwise to
the wall-normal and the span-wise components of the velocity during
the relaxation time (that $\propto 1/\Gamma$). Also, the model
describes well the part (about $50\%$ in the outer layer) of the total
kinetic energy that contains the streamwise components.

In the core of the flow ($\yp>450$) the model gives smaller values of 
all the components $\Wp_{ii}$, as compared to simulations and 
experiments. This is again because the model neglects the energy transfer 
toward the centerline of the channel, where the energy input into 
turbulence, $ -2 S W_{xy}$, disappears due to the symmetry reason.  

Also, there is a quantitative disagreement between the model and the simulations in 
the buffer layer.  One can relate this with the fact that the model 
neglects the energy flux toward the wall, which plays a considerable  role 
in the energy balance. The minimal models are local in space, but this 
effect can be effectively accounted for by an appropriate choice of the 
dissipation constants, taking $a_{yy}> a_{xx}=a_{zz}$. We do not propose
to take this route; in the buffer layer the 
turbulent flow  is strongly affected by highly intermittent events
(coherent structures) connected  with the near-wall instabilities of
the laminal sub-layer. This  is confirmed by the very large values
of the flatness (above 30), as  shown in  Fig.~\ref{f:flatness}.
Only for $\yp>50$ the flatness  reaches
the Gaussian value of 3 and one can successfully utilize various
lower-order closure model for describing wall bounded flows.

\section{\label{s:sum}Summary: strength and limitations of the minimal model}
The minimal model as formulated in this paper is a version of the algebraic Reynolds stress models. Its aim is to \textbf{\textbf{describe}}, for wall bounded turbulent flows,  the profile of mean flow and the statistics of turbulence on the level of simultaneous, one-point, second-order velocity correlation functions, i.e. the entries of the Reynolds-stress tensor $W_{ij}$.  The model was developed explicitly for plain geometry, including a wide variety of turbulent flows, like channel and plain  Couette flows,  to some extent fluid flows over inclined planes under gravity (modelling river flows), atmospheric turbulent boundary layers over flat planes and, in the limit of large Reynolds numbers, many other turbulent flows, including pipe, circular Couette flows, \emph{etc}. 

In developing a simple model one needs to decide what are the physically important aspects of the flow statistics, those which determine the mean-flow and the turbulent transport phenomena.
The choice of the Reynolds-stress approach was dictated by the decision to emphasize the accurate description of   $V(y)$- and $W_{ij}(y)$- profiles.   The main criteria in constructing the model  were simplicity, physical transparency, and maximal analytical tractability of the resulting model.  That is why we took liberty to ignore the spatial energy flux,  and, thanks to the plain geometry, to estimate the spatial derivatives and the outer scale of turbulence using the distance to the wall $y$.  The same motivations led to choosing the simplest  linear Rotta approximation of the ``Return to isotropy term"~\cite{51Rot} and the simplest  dimensional form of the nonlinear term for energy flux down the scales, also in agreement with ~\cite{51Rot}. 

By proper parametrization the number of fit parameters was reduced from twelve to four. Two of these, $a$ and $\~a$ are responsible for the viscous dissipation of the diagonal, $W_{ii}$ , and the off-diagonal, $W_{xy}$, components of the Reynolds-stress tensor. The other  two parameters - $b$ and $\~ b$ -- control the nonlinear relaxation of $W_{ii}$ and  $W_{xy}$. It appears that one cannot  decrease  the number of fit parameters further with impunity. The outer layer parameters $b=0.256$ and $\~ b=0.500$ where chosen to describe the observed constant values of \vK  in the log-law~\Ref{K-prof} and the asymptotic level of the density of kinetic energy. The viscous layer parameters were chosen to describe the observed values of the intersection $C$ in the \vK  log-law~\Ref{K-prof} and the peak of the kinetic energy in the buffer sub-layer. The resulting set of 5 equations for the mean shear $S(y)$, Reynolds stress $W_{xy}$ and $W_{xx}$, $W_{yy}$, $W_{zz}$ with just four fit parameters is referred to as the minimal model. 

As demonstrated in Sec.~\ref{s:exp} the minimal model with the given set~\Ref{consts} of  four parameters   describes five functions: 
\begin{itemize}
\item the mean velocity profile $V(y)$  is describe  with accuracy of $\simeq 1\%$ -- almost throughout the channel (except of small velocity defect in the core of the flow), cf.  Fig.~\ref{f:comp-meanV};  
\item
  the Reynolds stress profile $W_{xy}(y)$ is also  described with accuracy of  few percents (except in the viscous layer   $\yp<5$ in which $W_{xy}$ does not contribute to the mechanical balance), cf.  Fig.~\ref{f:comp-Rx};
\item
 the total kinetic energy profile $\frac12 W(y)$ is reproduced with reasonable (semi-quantitative) accuracy,  including the position and width of its peak in the buffer sub-layer, cf.  Fig.~\ref{f:4};
\item The profiles of the partial kinetic energies, $\frac12 W_{xx}(y)$, $\frac12 W_{yy}(y)$ and $\frac12 W_{zz}(y)$,  are
 reproduced, see Fig.~\ref{f:comp-Rii},  including the simple $\,\frac12\,$-$\,\frac14\,$-$\,\frac14\,$ distribution in the asymptotic outer region. This distribution is supported by recent experimental, DNS and LES data, as shown in Figs.~\ref{f:DNS-RelRii}, \ref{f:LES}, and  \ref{f:exp}.
\end{itemize}
   We consider all this as a good support of the minimal model;
 too much data is being reproduced to be an accident. It appears that the minimal model takes into account the essential physics  almost throughout the channel flow.  

On the other hand, one should accept that such a simple model  cannot pretend to describe all the aspects of the turbulent statistics in wall bounded flows. For example, the minimal model ignores the quasi-two dimensional character of turbulence and the existence of coherent structures in the very vicinity of the wall. The minimal model does not attempt to take into account many-point and high-order turbulent statistics, including three-point velocity correlation functions and pressure-velocity correlations, responsible for the spatial energy flux and for the isotropization of turbulence.  Finally, our choice of dissipation term definitely contradicts to the near-wall expansion,  (and see  Sec. 11.7.4 of  ~\cite{00Pope}),  in disagreement with various known  improvements of ~\cite{51Rot}. We propose that all this is a reasonable price for the simplicity and transparency of the minimal model, which is constructed with emphasis on the fundamental characteristics $V(y)$ and $W_{ij}(y)$ which are crucial for most applications. 

 We trust that a proper generalization of the minimal model will be found useful in the futures in studies of more complicated turbulent flows, laden with heavy particles, bubbles, \emph{\emph{etc}}.

\acknowledgements
We thank T.S. Lo for his critical reading of the manuscript and many insightful remarks.
We are grateful to Carlo Casciola for sharing with us his LES data. 
We express our appreciation to R. G.Moser, J. Kim, and N. N.Mansour, 
for making their comprehensive DNS data of high Re channel flow
available to all in Ref.~\cite{DNS}. This work was supported in part by the US-Israel
Binational Scientific Foundation and the European Commission under a TMR research grant.   SSZ acknowledges supports form the EU Marie Curie Chair Project MEXC-CT-2003-509742; ARO Project "Advanced parameterization and modelling of turbulent atmospheric boundary layers" - contract number W911NF-05-1-0055; and EU Project FUMAPEX EVK4-2001-00281.

\begin{figure}
 \centering\includegraphics[width=0.48 \textwidth]{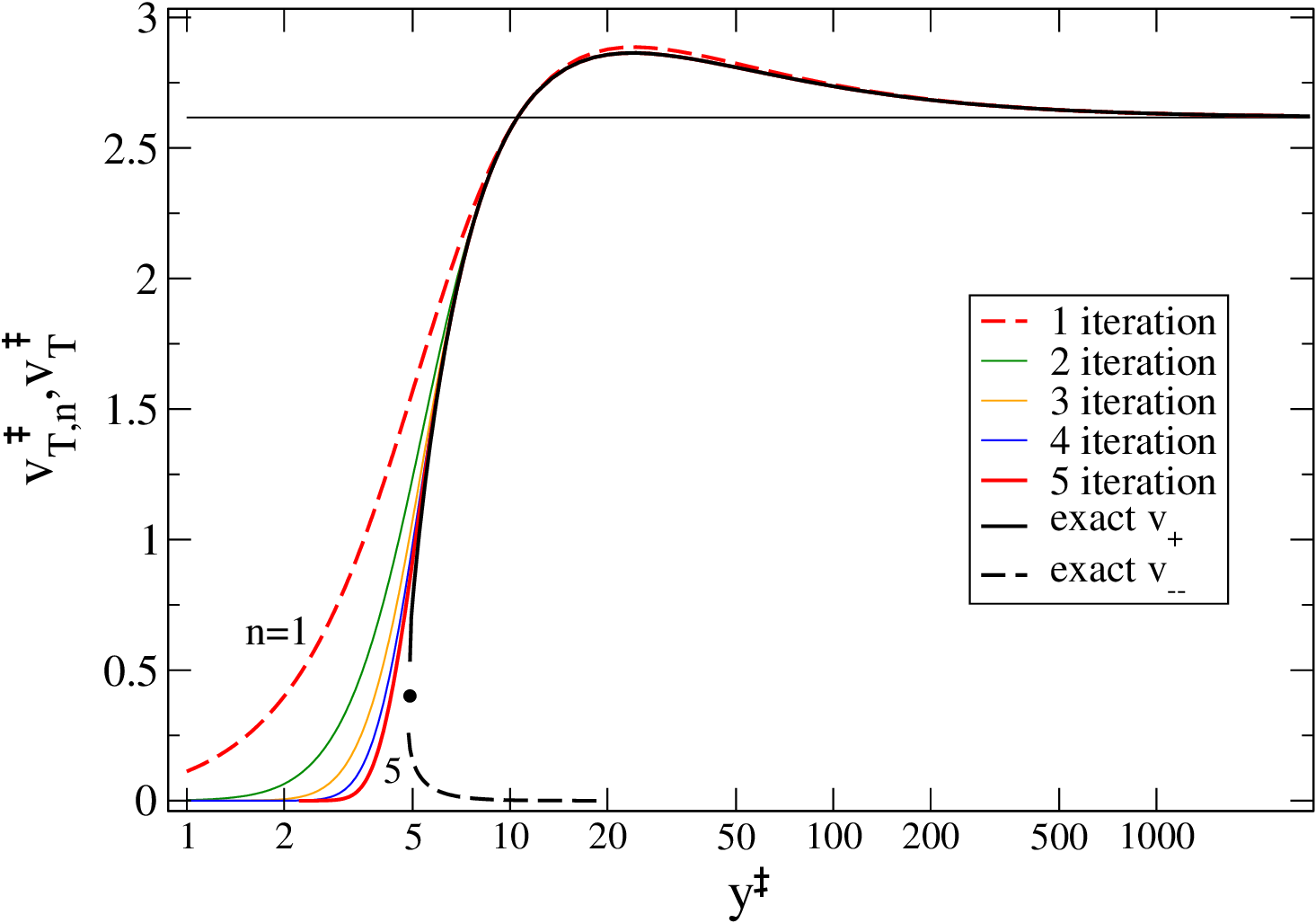}
\caption{\label{f:vt-n}  Log-plots of functions $\vd_\infty\approx
2.62$ (horizontal straight line), and profiles  $\vd_{n}(\yd)$ in
the sum-MM for $n=1$ (the dashed line), $2,\,3,\,4,\,5$ together with
the ``exact" (numerical) solutions of \REF{turb1d}, $\vd_+(\yd)$ (the
thick solid line) and $\vd_-(\yd)$ (dot-dashed line). Constants are
given by~\Ref{consts}.   For this choice the critical point
(designated as a black circle) corresponds to   $\vd_*\approx 0.4$ at $\yd\sb{vs}\approx 4.9$.}
\end{figure}

\appendix
\section{\label{a:iter}Validation of the iterative procedure} To see, 
how the iterative procedure described in Sec.~\ref{sss:iter} works, 
we  plotted in Fig.~\ref{f:vt-n}  iterative profiles of the turbulent 
velocity $\vd_{n}(\yd)$  for $n=1,\dots 5$ together with the 
(numerical) solutions of \REF{turb1d}, $\vd_+(\yd)$ (the thick solid line) 
and $\vd_-(\yd)$ (dot-dashed curve). The horizontal straight line  presents 
the asymptotic value $\vd_\infty$.   The critical point   $\{ \vd_*,\, 
\yd\sb{vs}\}$ is shown by a black circle. Our analysis shows (and see also 
Fig.~\ref{f:vt-n}), that already the simple Eq.~\Ref{iter3a} gives the 
relative accuracy (with respect to $\vd_\infty$) better than 1\% for 
$\yd>30$. The second iteration works with this accuracy in wider 
region  $\yd>10$, the third iteration gives 1\% accuracy for $\yd\approx 
5 $, which is about   the critical value $\yd\sb{vs}\approx 4.8$.   
Unexpectedly, the approximate solutions work even below the 
$\yd\sb{vs}$, where exact solution is $\vd=0$.  One observes with 
increasing $n$ the widening of  the region, in which $\vd_{n}$ 
practically indistinguishable from zero. The overall conclusion from 
these observations is that already the fist few iterations give a very 
good accuracy for all practical purposes, and very often one can use 
only the first or the second iteration.

\end{document}